\documentclass[twocolumn,floats,floatfix,showpacs,prd,superscriptaddress,nofootinbib]{revtex4-1}
\pdfoutput=1
\usepackage{amssymb,amsmath,amsthm,amsfonts}
\usepackage{bm} 

\usepackage[inline]{enumitem}
\usepackage[linktocpage]{hyperref}
\usepackage{url}
\usepackage[inline]{enumitem}

\usepackage{amsmath}
\usepackage{amsfonts}
\usepackage{amssymb}
\usepackage{dsfont}
\usepackage{amsthm}

\usepackage{stackrel,amssymb,amsmath}

\def\bPhi{{\bar{\Phi}}}
\def\gb{{\bar{g}}}
\def\cO{{\cal O}}
\def\cL{{\cal L}}
\def\cA{{\cal A}}

\begin{document}
		\title{Critical reflections on asymptotically safe gravity}
		
	\author{Alfio Bonanno}
	\affiliation{INAF, Osservatorio Astrofisico di Catania, via S.~Sofia 78 and INFN, Sezione di Catania, via S.~Sofia 64, I-95123,Catania, Italy}
	\author{Astrid Eichhorn}
	\affiliation{CP3-Origins, University of Southern Denmark, Campusvej 55, 5230 Odense M, Denmark}
	\affiliation{Institute for Theoretical Physics, Heidelberg University, Philosophenweg 16, D-69120 Heidelberg, Germany}
	\author{Holger Gies}
	\affiliation{Theoretisch-Physikalisches Institut, Abbe Center of Photonics,
		Friedrich-Schiller-Universit\"at Jena, and Helmholtz Institute Jena, Max-Wien-Platz 1, D-07743 Jena, Germany}
	\author{Jan M.~Pawlowski}
	\affiliation{Institute for Theoretical Physics, Heidelberg University, Philosophenweg 16, D-69120 Heidelberg, Germany}
	\author{Roberto Percacci}
	\affiliation{SISSA, via Bonomea 265, I-34136 Trieste and INFN, Sezione di Trieste, Italy
	}
	\author{Martin Reuter}
	\affiliation{Institute of Physics (THEP), University of Mainz, Staudingerweg 7, D-55099 Mainz, Germany}
	\author{Frank Saueressig}
	\affiliation{Institute for Mathematics, Astrophysics and Particle Physics (IMAPP), \\ Radboud University, Heyendaalseweg 135, 6525 AJ Nijmegen, The Netherlands}
	\author{Gian Paolo Vacca}
	\affiliation{INFN, Sezione di Bologna, Via Irnerio 46, I-40126 Bologna}

	\begin{abstract}
		Asymptotic safety is a theoretical proposal for the ultraviolet completion of quantum field theories, in particular for quantum gravity. Significant progress on this program has led to a first characterization of the Reuter fixed point. Further advancement in our understanding of the nature of quantum spacetime requires addressing a number of open questions and challenges. 
Here, we aim at providing a critical reflection on the state of the art in the asymptotic safety program, specifying and elaborating on open questions of both technical and conceptual nature. We also point out systematic pathways, in various stages of practical implementation, towards answering them. Finally, we also take the opportunity to clarify some common misunderstandings regarding the program. 
	\end{abstract}

\maketitle

\tableofcontents 
\vfill 
\eject

\section{Introduction and conclusions}
Asymptotic Safety \cite{Nink:2013fkw,Percacci:2017fkn,Reuter:2019byg} is a candidate for a quantum theory of the gravitational interactions. It does not require physics beyond the framework of relativistic Quantum Field Theory (QFT) nor does it require fields beyond the metric to describe the quantum geometry of spacetime.
Moreover, the inclusion of matter degrees of freedom, like
the standard model or its extensions, is conceptually straightforward.  Thus, ultimately, Asymptotic Safety may develop into a quantum theory comprising all fundamental  fields and their interactions.

The core idea of Asymptotic Safety was formulated by Weinberg \cite{Weinberg:1976xy,Weinberg:1980gg} in the late seventies. It builds on the insight of Wilson \cite{Wilson:1973jj}, linking the renormalizability and predictive power of a quantum field theory to fixed points of its Renormalization Group (RG) flow: a theory whose ultraviolet (UV) behavior is controlled by an RG fixed point does not suffer from unphysical UV divergences in physical processes like scattering events. The prototypical example for such a behavior is Quantum Chromodynamics (QCD) where the UV completion is provided by the free theory. In technical terms QCD is asymptotically free with the UV completion provided by a Gaussian fixed point.\footnote{The terminology Gaussian fixed point reflects that the action associated with the fixed point does not contain interactions and is thus quadratic in the fields.} It was then stressed in \cite{Weinberg:1980gg} that a valid UV completion could also be obtained from fixed points corresponding to actions with non-vanishing interactions, so-called non-Gaussian fixed points. In order to contrast this situation to asymptotic freedom, this non-trivial generalization has been termed ``asymptotic safety". Remarkably, the space of diffeomorphism invariant actions constructed from a four-dimensional (Euclidean) spacetime metric indeed seems to contain a non-Gaussian fixed point suitable for Asymptotic Safety, the so-called Reuter fixed point \cite{Reuter:1996cp,Souma:1999at}. 
	
As in other approaches to quantum gravity,
substantial progress has brought the program to a point
where a fair-minded assessment of its achievements
and shortcomings will be useful.
Therefore, the purpose of this article is to provide a
critical review of the current status of the field, of the
key open questions and challenges, and to point out
directions for future research.
By necessity, the discussion also covers questions of a more technical nature which is reflected in the character of some of the sections. This also entails that the article does not serve as an introduction to the asymptotic safety program, for which we refer the reader to the textbooks \cite{Percacci:2017fkn,Reuter:2019byg} and reviews \cite{Niedermaier:2006wt, Litim:2011cp, Reuter:2012id, Ashtekar:2014kba, Eichhorn:2017egq, Eichhorn:2018yfc, Reichert:2020mja}. A list of key references related to the open questions is provided within each section, pointing the reader towards the broader discussion in the literature.

The rest of the paper is organized as follows. In Sec.~\ref{sec:ASmainidea} we start with a concise introduction to asymptotic safety, also giving examples of non-Gaussian fixed points providing a UV completion in non-gravitational settings. The subsequent sections critically review open questions  along the following lines:
\begin{enumerate}[leftmargin=0.5cm]
\item Issues related to the use of the functional RG (FRG) (``uncontrollable approximations'', use of the background field method) are discussed in Sec.~\ref{FRG}.
\item Because of these theoretical uncertainties, it is important to cross-check the results with different methods. This is discussed in Sec.~\ref{sec:OtherApproaches}.
\item The difficulty of computing observables, and comparing with
observations, is discussed in Sec.~\ref{sec:Observables}.
\item Closely related to this is the, partly semantic, issue of the physical meaning of running couplings (can $\Lambda$ and $G$ run? If so, what are the physical implications of this running?)
and other aspects where the literature on asymptotic safety deviates
from standard particle physics procedures
(power vs. log running, use of dimensional regularization).
These points are discussed in Sec.~\ref{sec:runningcouplings}.
\item In Sec.~ \ref{sec:AS-EFT} we discuss whether and in what 
way asymptotic safety could be matched to effective field theory (EFT) at low energy. Here we also discuss
the limitations of the procedure of ``RG improvement''.
\item In Sec.~\ref{sec:ScaleA+Conformal} we address the relation between scale symmetry and conformal symmetry and the FRG. 
(How can one have scale invariance in the presence of $G$?)
We also critically review the argument that the entropy of black holes is incompatible with gravity being described by Asymptotic Safety
(``Gravity cannot be Wilsonian'' or ``Gravity cannot be a conformal field theory'').
\item The unsolved issue of unitarity is 
discussed in Sec.~\ref{sec:Unitarity}
(in particular: do higher derivatives imply ghosts?).
\item Finally, we stress the need of calculations in Lorentzian signature 
in Sec.~\ref{sec:LorentzianGravity}.
\end{enumerate}
The goal of this paper is threefold: 
\begin{itemize}[leftmargin=0.5cm]
\item[i)] reinforcing progress in the research field by clearly spelling out key open questions,
\item[ii)] strengthening a critical and constructive dialogue on asymptotically safe gravity within a larger community,
\item[iii)] contributing to a broad and critical assessment of the current status and future prospects of research avenues in quantum gravity.
\end{itemize}

	\section{Asymptotic Safety }\label{sec:ASmainidea}
		
	\subsection{The main idea}\label{sec:MainIdea}
	\emph{...where we recall the notion of quantum scale invariance and the predictive power of RG fixed points. }\\
	
Asymptotic Safety \cite{Percacci:2017fkn,Reuter:2019byg} builds on Wilson's modern view of renormalization, which links the renormalizability and predictive power of a quantum field theory to fixed points of its RG flow.\footnote{Generically, a fixed point will be neither UV nor IR, since it typically has both IR attractive (irrelevant) and IR repulsive (relevant) directions. Depending on the choice of RG trajectory, the fixed point can therefore induce a UV or an IR scaling regime. Given \emph{two} fixed points connected by an RG trajectory, the direction of the flow between them is fixed and the designation of UV and IR fixed point becomes unambiguous.}  
It is equivalent to the notions of ``quantum scale invariance in the UV''
and also to ``non-perturbative renormalizability'',
resulting in a theory that is fully specified by only a finite number of free parameters.
	
In practice, asymptotic safety is studied in the following way. One has a functional of the fields, that could be either a Wilsonian action $S_\Lambda$ depending on a UV cutoff $\Lambda$ or a generating functional $\Gamma_k$ for the one-particle irreducible (1PI) correlation functions depending on an IR cutoff $k$. We shall focus on the latter for definiteness, but at this stage the discussion is more general. For the present purposes, let us assume that this functional can be expanded in a suitable basis of operators $\{\mathcal{O}_i\}$, integrals of monomials in the field and its derivatives
\begin{align}
\label{eq:operatorexpansion}
\Gamma_k=\sum_i \bar{u}_i(k) \mathcal{O}_i. 
\end{align}
The beta functions of the, generally dimensionful, couplings $\bar{u}_i(k)$ are given by the derivatives of $\bar u_i(k)$ with respect to $t=\log k$. Then, one converts the dimensionful couplings\footnote{The notation $u_i$ for dimensionful quantities and $\tilde{u}_i$ for dimensionless quantities can also sometimes be found in the literature.}  $\bar{u}_i(k)$ into dimensionless ones by a suitable rescaling with the coarse-graining scale $k$,
\begin{equation}
u_i \equiv \bar{u}_i k^{-d_i} \, , 
\end{equation}
where $d_i$ is the canonical mass dimension of $\bar{u}_i(k)$. In this way one obtains a coupled set of autonomous differential equations
\begin{equation}\label{beta1}
k \partial_k	u_i(k)=\beta_{u_i}( \{ u_j	\} ) \, . 
\end{equation}
The solutions of this system are the RG trajectories
and each trajectory corresponds to a single physical theory. In general, it may happen that physical observables
diverge along a trajectory as $k\to\infty$ (e.g., at a Landau pole).
One simple way to avoid this is to require that the 
trajectory describing the physical world emanates from a fixed point as  $k$ is lowered from the UV to the IR.
At a fixed point $\{ u_{j*}	\}$ all beta functions vanish simultaneously, $\beta_{u_i}( \{ u_{j*}	\} ) = 0, \, \forall i$ and, as we shall discuss in more detail in Sec.~\ref{sec:ScaleA+Conformal}, scale invariance is realized\footnote{In most  cases this also implies conformal invariance.}. Such RG trajectories are said to be either asymptotically free or asymptotically safe theories. This should be contrasted to the case where physical observables blow up at a finite value of $k$ which indicates that one deals with an effective field theory.
	
The predictive power of asymptotic safety originates from the properties of the fixed point.
Linearizing the beta functions \eqref{beta1} about the fixed point,  and diagonalizing the stability matrix ${\bf B}_{ij} \equiv \partial_{u_j} \beta_{u_i} |_{u = u_*}$,
one can determine which directions are attractive and which ones are repulsive. 
Eigenvalues with positive (negative) real parts correspond to eigenvectors along which the flow (from UV to IR) is dragged towards (repelled by) the fixed point. One typically works with the scaling exponents\footnote{Note that the opposite sign convention, where the $\theta$ are defined without the additional negative sign, is also sometimes used in the literature.} $\theta_I = - {\rm eig}\, \bf{B}$.
Every irrelevant (IR attractive/UV repulsive/$\theta_I<0$) direction fixes one parameter in the initial conditions\footnote{More precisely, the ``memory" of the initial condition for an irrelevant direction is washed out by the RG flow and plays no role for the physics at $k=0$.} for $\Gamma_k$, whereas relevant (IR repulsive/UV attractive/$\theta_I>0$) directions correspond to free parameters. 
Marginal directions ($\theta_I=0$) typically only occur at Gaussian fixed points.
Thus, the number of independent free parameters of an asymptotically safe theory is equal to the number of relevant directions of the fixed
point that it originates from in the UV. 
At a free (Gaussian) fixed point, the relevant directions
correspond to couplings with positive mass dimension.
In a local theory, there is only a finite number of such parameters.
In principle, an interacting fixed point could have even
fewer relevant directions, and hence greater predictive power.
If one could integrate the RG flow to the IR, one could test if the 
low-energy relations implied by these properties of the UV fixed point are verified or not,
cf.~Sec.~\ref{sec:Observables} for further discussion.

\subsection{Non-gravitational examples}
\emph{... where we provide a list of non-gravitational, asymptotically safe theories together with the corresponding mechanism for asymptotic safety and we discuss how several techniques are used to study these examples.}\\
	
Whereas the existence of UV-complete quantum field theories based on
the mechanism of asymptotic safety has been anticipated already in 
the early days of the RG \cite{GellMann:1954fq,Johnson:1967pk}, concrete examples have
been identified only much later, as a parametric control beyond perturbation
theory is typically required. 
A paradigmatic class of examples is given by fermionic
models in $d=3$ dimensional spacetime including, for instance, the Gross-Neveu model: though interactions of the type
$\sim (\bar\psi \mathfrak{m}\psi)^2$ (with $\mathfrak{m}$ carrying some internal
spin and/or flavor structure) belong to the class of perturbatively
non-renormalizable models, there is by now convincing evidence that a large
class of such models are in fact asymptotically safe in $2<d<4$ dimensional
spacetime. Initially, the existence of the underlying non-Gaussian fixed
points has been demonstrated by means of $1/N$ expansions
\cite{Parisi:1975im,Gawedzki:1985ed}; indeed, non-perturbative
renormalizability has been proved for specific models to all orders in the
$1/N$ expansion \cite{deCalan:1991km} with explicit results for higher orders
being worked out, e.g., in
\cite{Gracey:1990sx,Vasiliev:1992wr,Hands:1994kb,Gracey:2016mio}. Further quantitative evidence subsequently came from
$2+\epsilon$ or $4-\epsilon$ expansions \cite{Hikami:1976at,Rosenstein:1988pt,Gat:1991bf,Rosenstein:1993zf,Mihaila:2017ble,Zerf:2017zqi}; the FRG for the
first time facilitated analytic computations directly in $d=3$
\cite{PhysRevLett.86.958,Hofling:2002hj,Braun:2010tt,Gies:2010st,Gehring:2015vja,Classen:2015mar,Vacca:2015nta,Knorr:2016sfs}. For
the asymptotic safety
program, these models are instructive for several reasons:
\begin{itemize}[leftmargin=0.5cm]
\item[\textit{(i)}] The fermionic non-Gaussian fixed point is typically connected to
a quantum phase transition. The latter is characterized by universal
critical exponents which can also be studied using simulational methods
\cite{Hands:1992be,Karkkainen:1993ef,Christofi:2006zt,Chandrasekharan:2011mn,Chandrasekharan:2013aya,Wang:2014cbw,Hesselmann:2016tvh,Li:2014aoa,Hands:2016foa,Schmidt:2016rtz,Lenz:2019qwu}
or the conformal bootstrap \cite{Vasiliev:1992wr,Iliesiu:2017nrv}. In
this way, the variety of available approaches have led to a confirmation of asymptotic safety of these models to
a substantial degree of quantitative precision, summarized, e.g., in \cite{Ihrig:2018hho}.
\item[\textit{(ii)}] While analytical as well as path integral Monte Carlo
computations are typically performed in Euclidean spacetime, these models
are relevant for layered condensed-matter ``Dirac materials'' \cite{Wehling:2014cla,Vafek:2013mpa},
corresponding to a $d=2+1$ dimensional spacetime with Lorentzian
signature. The quantitative agreement also with Quantum Monte Carlo methods (based on a Hamiltonian formulation) \cite{Wang:2014cbw,Hesselmann:2016tvh,Li:2014aoa}, demonstrates that
asymptotic safety of these systems is visible in Euclidean as well as
Lorentzian formulations.
\item[\textit{(iii)}] As a generic mechanism of asymptotic safety in these
		models, an irrelevant (i.e., perturbatively non-renormalizable) operator
		such as the fermionic interaction $\sim (\bar\psi \mathfrak{m}\psi)^2$
		becomes relevant as a consequence of strong fluctuations. Correspondingly,
		the anomalous dimension of this and subsequent operators is shifted by an
		amount of $\mathcal{O}(1)$; see, e.g., \cite{Braun:2010tt,Jakovac:2014lqa} for a
		determination of an infinite set of scaling dimensions for large $N$. As a
		consequence, strongly power-counting irrelevant operators remain irrelevant
		and do not introduce an unlimited set of new physical parameters. The same
		pattern is also observed in many studies of asymptotically safe gravity
		\cite{Codello:2008vh,Benedetti:2009rx,Falls:2013bv,Falls:2014tra,Eichhorn:2015bna,Alkofer:2018fxj}.
		\item[\textit{(iv)}] The comparative simplicity of these models has enabled a
		first study of the momentum dependence of 4-point correlation functions at
		the non-Gaussian fixed point \cite{Dabelow:2019sty}. For instance, the Gross-Neveu model
		($\mathfrak{m}=\mathds{1}$) in $d=2+1$ at the non-Gaussian fixed point
		can be analyzed in terms of an $s$-channel-dependent Gross-Neveu coupling
		$g_\ast(s)$ which depends nontrivially and non-analytically on the
		dimensionless $s$ variable at the UV fixed point. In fact, the $s$ channel
		dependence can be shown to dominate over possible $t$ and $u$ channel
		dependences in a quantifiable manner at large $N$, resulting in a simpler form factor-like structure of the 1PI 4-vertex at the UV fixed point.
		This illustrates that scattering properties in the scaling regime can develop
		nontrivial features beyond the scaling suggested by naive power-counting.
		
	\end{itemize}
	
Further examples for asymptotic safety include Yang-Mills theory in
$d=4+\epsilon$ \cite{Peskin:1980ay,Kazakov:2002jd,Gies:2003ic,Morris:2004mg},
and non-linear sigma models in $d=2+\epsilon$
\cite{Polyakov:1975rr,Bardeen:1976zh,Friedan:1980jf,Higashijima:2003rp,Codello:2008qq,Fabbrichesi:2010xy};
for the latter, there is clear evidence for asymptotic safety even in $d=3$
from lattice simulations \cite{Wellegehausen:2014noa}. The limit of large number of fermions $N_f$ in gauge theories has recently seen a resurgence of interest, e.g., \cite{Antipin:2017ebo, Antipin:2018zdg,Dondi:2020qfj, Dondi:2019ivp}, with early work in \cite{PalanquesMestre:1983zy,Gracey:1996he}, see also \cite{Holdom:2010qs}.
	
Another recently discovered set of asymptotically safe models is given by gauged Yukawa
models in the Veneziano limit of a suitably arranged large number of vector
fermions  $N_f$  adjusted to the number of colors $N_c$ of the gauge group
\cite{Litim:2014uca,Bond:2016dvk,Bond:2017lnq,Bajc:2016efj,Mann:2017wzh,Bond:2017wut,Pelaggi:2017abg}
in $d=3+1$ dimensional spacetime. 
Contrary to the lower-dimensional fermionic
models, these gauged Yukawa models are power-counting renormalizable to all
orders in perturbation theory. Because of the large number of fermions, fermionic screening dominates the running of the gauge coupling, such that asymptotic freedom is lost. The RG flow at high energies nevertheless remains bounded, as it is controlled by a UV fixed point appearing in all RG marginal couplings. Whereas
perturbative renormalizability of these models supports the use of
perturbative RG beta functions in the first place, the existence of
non-Gaussian UV fixed points is parametrically controlled by a suitably
small Veneziano parameter, e.g., $\epsilon=\frac{N_{\text{f}}}{N_{\text{c}}}-
\frac{11}{2}$ as in \cite{Litim:2014uca}. Despite this technical vicinity to
perturbative computations, the behavior of the theory near the fixed point is
very different from the perturbative behavior near the Gaussian fixed
point. For instance, the perturbatively marginal operators turn into
(ir-)relevant operators with anomalous dimensions reaching up to
$\mathcal{O}(1)$ for $\epsilon \lesssim \mathcal{O}(0.1)$. The couplings
therefore scale with a power of the RG scale rather than
logarithmically. Also, higher-order operators -- though remaining RG
irrelevant -- generically acquire non-trivial fixed-point values and can thus
exert an influence on scattering properties at highest energies.

\section[Functional renormalization group]{Functional Renormalization Group}\label{FRG}

In Section~\ref{sec:ASmainidea} we have discussed the asymptotic-safety mechanism without referring to any specific calculation method. Now we introduce the Functional Renormalization Group (FRG), which has been the main tool enabling progress in Asymptotic Safety in the last 20 years. It has been successfully applied to a large number of other theories and physical phenomena, in particular non-perturbative ones. Applications range from the phase structure of condensed matter systems, to confinement and chiral symmetry breaking in QCD, to the electroweak phase transition in the early universe and beyond Standard Model physics. In cases, where results from other non-perturbative methods 
(lattice simulations, Dyson-Schwinger equations, Resurgence etc.) exist, the FRG results compare well to those obtained by other methods. It is also worth emphasizing that, while the combination of conceptual and technical challenges 
in quantum gravity is certainly unique, many of the technical challenges and physical effects encountered here have counterparts in other theories, most notably in non-Abelian gauge theories, where they can also be tested against other non-perturbative methods.   

\subsection{Brief introduction to the FRG}\label{sec:FRG+Intro}
\emph{...where we briefly introduce the FRG as a tool to calculate the effective action.}\\

Currently, the primary tool to investigate Asymptotic Safety is the Functional Renormalization Group (FRG) equation for the effective average action $\Gamma_k$ introduced in \cite{Wetterich:1992yh, Ellwanger:1993mw, Morris:1993qb} (Wetterich equation), and in \cite{Reuter:1996cp} for gravity. $\Gamma_k$ depends on the content of the theory at hand, in quantum gravity it contains the metric degrees of freedom, Faddeev-Popov ghosts 
and possibly also matter fields. In the FRG approach the scale $k$ is an infrared cutoff scale below which quantum fluctuations are suppressed. 
Thus, $\Gamma_k$ encodes the physics of quantum fluctuations above the cutoff scale.  
For $k\to 0$, all quantum fluctuations have been taken into account and $\Gamma_{k=0}$ is the full quantum effective action, 
\begin{align}
\Gamma = \lim_{k \rightarrow 0} \, \Gamma_k \, 
\end{align}
whose minimum is the vacuum state of the QFT.
The flow equation for $\Gamma_k$ encodes the response of the effective average action $\Gamma_k$ to the process of integrating out quantum fluctuations within a momentum shell, 
\begin{align}
\label{FRGE}
k \partial_k	\Gamma_k[\Phi;\bPhi]
=
\frac{1}{2}	\mathrm{Tr} \left[	\frac{1}{\Gamma_k^{(2)}[\Phi;\bPhi]+ R_k}	\, k \partial_k	R_k	\right]\,.
\end{align}
The term $(\Gamma_k^{(2)}+R_k)^{-1}$ on the right hand side of \eqref{FRGE} is the propagator in the regularized theory. Here we have introduced $\Gamma_k^{(2)}=\Gamma^{(\Phi\, \Phi)}$, the second derivative of $\Gamma_k$ w.r.t.\ the fields $\Phi$. In \eqref{FRGE} we have also introduced 
a  generic background $\bPhi$ which typically is chosen as the solution to the quantum equations of motion. Then, the fluctuation field 
$\Phi$ encodes the fluctuations about this background, and the 1PI correlation functions of the fluctuation fields $\langle \Phi_{i_1} \cdots \Phi_{i_n}\rangle_\textrm{1PI}$ (proper vertices) in a given background $\bPhi$ are given by     
\begin{equation}
\label{propervertices}
\Gamma^{(\Phi_{i_1}\cdots \Phi_{i_n})}_k[\bPhi] \equiv \left. \frac{\delta}{\delta \Phi_{i_1}} \cdots \frac{\delta}{\delta \Phi_{i_n}} \Gamma_k[\Phi;\bPhi]\right|_{\Phi = 0} \, .
\end{equation} 
The term $R_k$ is a cutoff scale $k$- and momentum-dependent infrared regulator which suppresses fluctuations with momenta $p^2  \lesssim k^2 $, decays rapidly for momenta $p^2 \gtrsim k^2$, and vanishes at $k^2=0$. The second property renders the flow equation \eqref{FRGE} finite due to the decay of $k\partial_k R_k$ for large momenta. The regulator $R_k$ is independent of the fluctuation field, but may carry a dependence on the background field. In a quantum field theory in flat space typically $p^2$ is the plain momentum squared, while in gravity and gauge theories $p^2$ may be associated with a background-covariant Laplacian. Finally, $\mathrm{Tr}$ comprises a sum over all fluctuation fields and an integral over (covariant) loop momenta. The corresponding loop integration is peaked about momenta $p^2 \approx k^2$, leading to the momentum-shell integration. In summary, the flow equation \eqref{FRGE} transforms the task of performing the path integral into the  task of solving a functional differential equation.

Conceptually, the Wetterich equation implements the idea of the Wilsonian Renormalization Group: lowering $k$ corresponds to integrating out  quantum fluctuations shell by shell in momentum space. For $k\to\infty$, the theory approaches the bare or renormalized ultraviolet action, depending on the underlying renormalization procedure, for a detailed analysis see, e.g.,~\cite{Pawlowski:2005xe, Manrique:2008zw, Manrique:2009tj, Morris:2015oca, Rosten:2010pc}. The fact that eq.\ \eqref{FRGE} does not require specifying a bare action a priori makes it a powerful tool to scan for (interacting) RG fixed points and study their properties. The bare action can then be reconstructed from the RG fixed point along the lines of \cite{Manrique:2008zw,Morris:2015oca}. Essentially, the Wetterich equation can be viewed as a tool to  systematically test which choice of bare action gives rise to a well-defined and predictive path integral for quantum gravity.

Notably, if one approximates $\Gamma_k^{(2)}$ by the $k$-independent second functional derivative of a given bare action $S^{(2)}$, one obtains 
\begin{equation}
\Gamma_k \approx S+\frac{1}{2} {\rm Tr} \log \left( S^{(2)}+R_k\right), \label{eq:tracelog}
\end{equation}
which reduces to the standard one-loop effective action for $k=0$.
Accordingly, approximations to the FRG always contain one-loop results in a natural way.

\subsection{FRG approach to quantum gravity}
\label{sec:FRG+Gravity}
\emph{...where we review the Functional Renormalization Group approach to quantum gravity, with a particular focus on background-field techniques.}\\

In the gravitational context, the construction of eq.\ \eqref{FRGE} makes use of the background field method, decomposing the physical metric $g_{\mu\nu}$ into a fixed, but arbitrary background metric $\gb_{\mu\nu}$ and fluctuations $h_{\mu\nu}$, see \cite{Niedermaier:2006wt} for technical details\footnote{Notably, the asymptotic-safety mechanism is not tied to the spacetime metric carrying the gravitational degrees of freedom. 
While explored in far less detail, the vielbein
and the Palatini formalisms may also lead to a theory which
is asymptotically safe \cite{Daum:2010qt,Harst:2012ni,Dona:2012am,Daum:2013fu,Harst:2014vca,Reuter:2015rta,Harst:2015eha}.}. The typical example is the linear split, 
\begin{equation}
\label{eq:LinearSplit}
g_{\mu\nu}=\bar g_{\mu\nu}+h_{\mu\nu} \, . 
\end{equation}
In the literature the fluctuation field $h_{\mu\nu}$ is commonly multiplied with the square root of the Newton constant which makes it a standard dimension-one tensor field in four spacetime dimensions. The linear split \eqref{eq:LinearSplit} is the common choice not only in quantum gravity but also in applications of the background field method to gauge theories or non-linear sigma models. In gravity it comes at the price that the fluctuation field $h_{\mu\nu}$ is not a metric field, indeed it has no geometrical meaning. While this is not necessary, alternative parameterizations  
have been used. These  
have the general form
\begin{equation}\label{eq:GenSplit}
g_{\mu\nu} = f(h,\bar g)_{\mu}{}^{\kappa}\bar{g}_{\kappa\nu}\,.
\end{equation}
Of these alternative cases, the exponential split with $f(h, \bar{g}) = \rm exp[\bar{g}^{-1}h]$ has been explored, e.g., in \cite{Nink:2014yya,Eichhorn:2015bna,Dona:2015tnf,Percacci:2015wwa,Labus:2015ska}. Further, the geometrical split in the Vilkovisky-deWitt approach with a diffeomorphism invariant flow has been studied in \cite{Branchina:2003ek, Pawlowski:2003sk, Pawlowski:2005xe, Donkin:2012ud, Demmel:2014hla}, for applications to non-linear sigma models see \cite{Safari:2016dwj, Safari:2016gtj}. 

Different parameterizations \eqref{eq:GenSplit} only constitute the same quantization if they (i) cover the same configuration space and (ii) the Jacobian that arises in the path integral is taken into account  (see \cite{Percacci:2015wwa} for a related discussion). Condition (i) does not hold, e.g., for linear parameterization and exponential parameterization, see, e.g., \cite{Nink:2014yya,Percacci:2015wwa}, while the linear split and the geometrical one with the Vilkovisky connection at least agree locally. However, it is well-known from two-dimensional gauge theories, that quantizations on the algebra and on the group 
can differ, see, e.g., \cite{Birmingham:1991ty}. Moreover, studies of the parameterization dependence of results in truncations, e.g.,  \cite{Gies:2015tca,Ohta:2016npm,Ohta:2016jvw,deBrito:2018jxt}, so far do not account for (ii).

The presence of the background allows to discriminate ``high-" and ``low-momentum" modes by, e.g., comparing their eigenvalues with respect to the background Laplacian to the coarse-graining scale $k$. Moreover, it also necessarily enters gauge-fixing terms for the fluctuation field. As a consequence, the effective action $\Gamma_k$ inherits two arguments, the set of fluctuation fields $\Phi$ and the corresponding background fields $\bar{\Phi}$ for all cutoff scales $k$. We emphasize that this also holds true for vanishing cutoff scale, $k=0$, due to the gauge fixing. 

Conceptually, the Wetterich equation lives on the so-called theory space, the space containing all action functionals constructable from the field content of the theory and compatible with its symmetry requirements. The FRG then defines a vector field generating the RG flow on this space. 
We proceed by discussing two systematic expansion schemes commonly used in quantum gravity (as well as other systems): the vertex expansion and the (covariant) derivative expansion.

The proper vertices of the effective average action \eqref{propervertices} can be used as coordinates in theory space as the set of 
(1PI) correlation functions $\{\Gamma_k^{(\Phi_{i_1}\cdots \Phi_{i_n})}[\bPhi]\}$ defines a given action and hence a theory. The vertex expansion is the expansion in the order of the fluctuation correlation functions and hence in powers of the fluctuation field,  
\begin{align}
\nonumber 
\Gamma_k[\Phi; \bar{\Phi}]=&\, \sum_{n=1}^\infty \frac{1}{n!}\int \prod_{i=1}^n \left[{\rm d}^d x_i  \, \Phi_{j_i}(x_i) \right] \times\\[1ex] 
&\, \qquad \times \Gamma_{k}^{(\Phi_{j_1}\cdots \Phi_{j_n})}[\bar\Phi](x_1,\cdots,x_n) 
\,,
\label{eq:EffActvertex}
\end{align}
where $\Gamma_k^{(\Phi_{j_1}\cdots \Phi_{j_n})}[\bar\Phi]$,
for $n>2$, are the proper vertices \eqref{propervertices}, that carry the measure factors $\prod_i \sqrt{ \bar{g}(x_i)}$. 

The derivative expansion is best explained in the case of the diffeomorphism invariant background effective action $\bar\Gamma_k[\bPhi]=\Gamma_k[\Phi=0;\bPhi]$. This object can be expanded in diffeomorphism invariant operators such as powers of the curvature scalar and other invariants. Then, in the derivative expansion the sum in
\eqref{eq:operatorexpansion} contains all diffeomorphism invariant terms with less than a certain number of derivatives. The leading order of this expansion is
\begin{equation}\label{eq:EffActgeo}
\bar\Gamma_k[g_{\mu\nu}] \simeq \frac{1}{16 \pi G_k} \int d^dx  \sqrt{g} \left[ 2 \Lambda_k - R \right] \, .  
\end{equation}
At the next order 
one has to add four-derivative terms including $R^2$, $R_{\mu\nu}R^{\mu\nu}$, and $R_{\mu\nu\rho\sigma}R^{\mu\nu\rho\sigma}$, and so on. In this light, it should be understood that the Einstein-Hilbert action just provides the leading terms in the derivative expansion of $\bar\Gamma_k[g_{\mu\nu}]$ and does not constitute the bare action underlying Asymptotic Safety.
It has to be supplemented by gauge-fixing and ghost terms,
and, if the approximation is extended,
additional terms $\widehat{\Gamma}_k[h;\gb]$ depending on two arguments separately.
``Bimetric" studies distinguishing $g_{\mu\nu}$ and $\bar{g}_{\mu\nu}$ for the Einstein-Hilbert truncation can be found in \cite{Manrique:2009uh,Manrique:2010am,Becker:2014qya}.

\subsection{Results for asymptotically safe gravity}
\label{sec:FRG+Results}
\emph{...where we give a brief overview of the results 
obtained with the truncated FRG 
and provide a sketch of the full flow from the UV fixed point down to the IR.}\\ 

Most work has been done in the background-field approximation, that is  
$\Gamma_k[\Phi;\bPhi]= \bar \Gamma_k[\Phi+\bPhi]+$\textit{gauge fixing + ghosts}. 
If one evaluates the FRG in a one-loop approximation, including terms quadratic in
curvature, the known universal beta functions of the four-derivative couplings
are reproduced, but additionally the cosmological and Newton constant have a
non-trivial fixed point \cite{Codello:2006in,Niedermaier:2009zz,Niedermaier:2010zz}. Going beyond one loop, the following classes of operators have been studied
in pure gravity: The Einstein-Hilbert truncation has been explored extensively \cite{Reuter:1996cp,Dou:1997fg,Souma:1999at,Lauscher:2001ya,Reuter:2001ag,Litim:2003vp,Groh:2010ta,Eichhorn:2010tb,Nagy:2013hka,Falls:2014zba,Falls:2015qga,Gies:2015tca}.
Einstein-Hilbert action plus $R^2$ \cite{Lauscher:2002sq,Rechenberger:2012pm}; Einstein-Hilbert action plus $R^2$ and $R_{\mu\nu\rho\sigma}R^{\mu\nu\rho\sigma}$ \cite{Benedetti:2009gn,Benedetti:2009rx,Ohta:2013uca,Ohta:2015zwa,Hamada:2017rvn}; 
Einstein-Hilbert action plus $R^2$, $R_{\mu\nu}R^{\mu\nu}$ and $R_{\mu\nu\rho\sigma}R^{\mu\nu\rho\sigma}$ \cite{Falls:2020qhj};
Einstein-Hilbert action plus the Goroff-Sagnotti counterterm $R^{\mu\nu}{}_{\rho\sigma}R^{\rho\sigma}{}_{\alpha\beta}R^{\alpha\beta}{}_{\mu\nu}$ \cite{Gies:2016con}; polynomial functions of the scalar curvature (polynomial ``$f(R)$ truncation'') up to orders $N=6$ \cite{Codello:2007bd,Machado:2007ea}, $N=8$ \cite{Codello:2008vh}, $N=35$ \cite{Falls:2013bv,Falls:2014tra}, and lately also $N=71$ \cite{Falls:2018ylp},
or effective actions of the form 
$f_1(R_{\mu\nu}R^{\mu\nu})+f_2(R_{\mu\nu}R^{\mu\nu}) R$,
where $f_1$ and $f_2$ are polynomials
\cite{Falls:2017lst}, 
effective actions of the form 
$f_1(R_{\mu\nu\rho\sigma}R^{\mu\nu\rho\sigma})+f_2(R_{\mu\nu\rho\sigma}R^{\mu\nu\rho\sigma}) R$
where $f_1$ and $f_2$ are polynomials 
or finally effective actions consisting of a single trace of
$n$ Ricci tensors ($R^\mu{}_\nu R^\nu{}_\rho\ldots R^\alpha{}_\mu$)
with $n$ up to 35, \cite{Kluth}. 
The case of an ``infinite number" of couplings has been addressed
in the $f(R)$ truncation by solving \cite{Demmel:2012ub,Dietz:2012ic,Dietz:2013sba,Demmel:2013myx,Demmel:2014sga, Demmel:2014hla,Demmel:2015oqa,Dietz:2015owa,Ohta:2015efa,Ohta:2015fcu,  Dietz:2016gzg, Morris:2016spn, Christiansen:2017bsy, Gonzalez-Martin:2017gza, Alkofer:2018baq, Burger:2019upn}  a non-linear differential equation for $f$ \cite{Codello:2007bd, Machado:2007ea,Benedetti:2012dx,Demmel:2012ub,Dietz:2012ic,Dietz:2013sba,Demmel:2013myx,Demmel:2014sga, Demmel:2014hla,Falls:2014tra,Demmel:2015oqa,Dietz:2015owa,Ohta:2015efa,Ohta:2015fcu,  Dietz:2016gzg, Falls:2016msz,Morris:2016spn, Christiansen:2017bsy, Gonzalez-Martin:2017gza,Falls:2017lst,deBrito:2018jxt, Falls:2018ylp, Alkofer:2018baq, Burger:2019upn}. 
 Global solutions for such ``infinite" truncations can also be found for gravity coupled to a scalar field, see, e.g.,  \cite{Borchardt:2015rxa}.
For a more general overview of the situation in gravity-matter systems we refer to the review \cite{Eichhorn:2018yfc}. Notably, a fixed point suitable for Asymptotic Safety has been identified in all these works.

As is clear from this list, the terms included do not reflect
the systematics of a derivative expansion.
It has also to be said that in many of these calculations the beta
functions that one obtains are only unknown linear combinations of the beta functions
that would be obtained if all curvature invariants of the same order were included.
This is because the calculations are done on spheres, e.g., \cite{Lauscher:2002sq,Rechenberger:2012pm,Codello:2007bd,Machado:2007ea,Codello:2008vh,Falls:2013bv,Falls:2014tra,Falls:2018ylp,Falls:2017lst,Benedetti:2012dx,Dietz:2013sba,Bridle:2013sra,Dietz:2012ic,Demmel:2013myx,Demmel:2015oqa,Dietz:2015owa,Ohta:2015efa,Ohta:2015fcu,Morris:2016spn,Gonzalez-Martin:2017gza,deBrito:2018jxt},  a hyperbolic background \cite{Falls:2016msz}
or sometimes on Einstein backgrounds, e.g., \cite{Benedetti:2009gn,Benedetti:2009rx}, and this does not
permit to differentiate between functions of Ricci tensor
and of the Ricci scalar, for example.

In terms of the vertex expansion, most work has been built on an expansion around flat space while keeping part of the full momentum dependence of propagators and vertices. For the vertices typically the symmetric point configuration is considered. For results in pure gravity and gravity-matter systems see  \cite{Christiansen:2012rx, Codello:2013fpa, Christiansen:2014raa, Meibohm:2015twa, Christiansen:2015rva, Denz:2016qks, Christiansen:2017cxa, Knorr:2017fus, Eichhorn:2018akn, Eichhorn:2018ydy,  Eichhorn:2018nda}. These works have revealed the existence of a nontrivial
fixed point in the two-, three-, and four-point functions compatible with the findings in the background approximations. Analogous calculations with compatible results have also been done for the two-and  
three-point functions on a spherical background \cite{Christiansen:2017bsy, Burger:2019upn}. The results in \cite{Christiansen:2017bsy,Burger:2019upn} for background curvature and background momentum-dependent two- and three-point function of the fluctuation field have then been used to compute the full $f(R)$-potential in pure gravity and in the gravity-scalar system beyond the background approximation. 

Just like in the derivative expansion in asymptotically safe gravity, it has also not been possible to fully and systematically
implement the vertex expansion beyond the lowest order: In particular, 
the three- and four-point functions have only been calculated 
for a special kinematical configuration and the symmetric background does not allow to fully disentangle different operators.

To connect the UV fixed point to physics at $k=0$, complete trajectories must be constructed. Currently, this part of the program is less advanced than the characterization of the fixed point itself; UV-IR flows have been computed, e.g.,  in \cite{Christiansen:2012rx, Donkin:2012ud, Christiansen:2014raa, Denz:2016qks, Knorr:2019atm}. It is expected that complete solutions are most likely characterized by several regimes \cite{Reuter:2001ag,Reuter:2004nx,Rechenberger:2012pm,Gubitosi:2018gsl}, see also, e.g., \cite{Eichhorn:2017ylw,Eichhorn:2019tcj} for matter-gravity systems:

\noindent
- The first part of the flow from the Reuter fixed point in the UV down to some scale $M_1$
is in a linear regime close to the fixed point. 
At these extreme UV scales, the system could {\it a priori} either be in a strongly interacting non-perturbative regime or 
be characterized by weak interactions. 
There are some tentative hints for the latter (see Sec.~\ref{sec:convergenttruncations}), but a conclusive statement regarding the nature of the fixed point cannot yet be made.

\noindent
- Close to the Planck scale, 
the flow has potentially already left the linear regime around the fixed point. 
In simple approximations, $M_1 = M_{\rm Pl}$, i.e., the transition scale at which fixed-point scaling stops, actually comes out equal to the Planck scale. 
The regime around the Planck scale could again be characterized by either non-perturbative or near-perturbative physics -- 
irrespective
of the nature of the fixed point. 
Once one leaves the fixed-point regime, 
non-localities of order $1/k$, or 
dynamically generated scales are expected to play a role.
\footnote{It is important to realize that non-local operators, i.e., operators with inverse powers of derivatives, proliferate under the flow and are canonically increasingly relevant. They are therefore likely to destroy the predictive nature of the fixed point, if included in theory space explicitly. On the other hand, the flow never generates an operator with inverse powers of derivatives within a quasi-local theory space, i.e., the requirement of quasilocality can be imposed consistently on the theory space. Of course it is well-known that the full effective action contains physically important non-localities. These arise in the limit $k \rightarrow 0$ and are expected to be captured through resummation of quasi-local operators, see, e.g., \cite{Codello:2010mj} for an example. This intricacy potentially makes this regime of the flow difficult to describe in a quantitatively robust way. It is generally expected that an expansion of the effective action in terms of vertex functions or form factors is best suited to this regime, as it can automatically capture non-localities of order $1/k$, and also encode the presence of dynamically generated scales.} \\
- 
Below the Planck scale, the description of the purely gravitational sector
is expected to become much simpler. 
Once near the Gaussian fixed point, the flow is dominated by the canonical scaling terms. For instance, the dimensionful Newton constant becomes scale independent.
One expects that corrections obtained within the effective-field theory approach to quantum gravity are recovered in this regime.

In many cases these works on asymptotic safety based on the FRG
can be compared to, or substantiated by, other approximation methods
or techniques.
We defer a discussion of such relations to Sects. 
\ref{sec:OtherApproaches} and \ref{sec:AS-EFT}.

\subsection{The convergence question}\label{sec:convergenttruncations}
\emph{...where we discuss the convergence (or lack thereof) of 
systematic expansion schemes in the FRG.}\newline

In practical applications, one has to work in truncations of the theory space. These can also be infinite dimensional, if a closed form for the flow of an appropriate functional can be found. In the gravitational case, closed flow equations for $f(R)$ truncations constitute an example \cite{Codello:2007bd, Machado:2007ea,Benedetti:2012dx,Demmel:2012ub,Dietz:2012ic,Dietz:2013sba,Demmel:2013myx,Demmel:2014sga, Demmel:2014hla,Falls:2014tra,Demmel:2015oqa,Dietz:2015owa,Ohta:2015efa,Ohta:2015fcu,  Dietz:2016gzg, Falls:2016msz,Morris:2016spn, Christiansen:2017bsy, Gonzalez-Martin:2017gza,Falls:2017lst,deBrito:2018jxt, Falls:2018ylp, Alkofer:2018baq, Burger:2019upn}.  Further examples are the scalar potential and a nonminimal functional in scalar-tensor theories, see, e.g., \cite{Narain:2009fy, Henz:2013oxa, Percacci:2015wwa, Labus:2015ska, Henz:2016aoh}.

A reasonable expansion scheme should capture the relevant physics already at low orders of the expansion. For a fixed point, this includes the relevant operators. At the free fixed point one simply expands according to canonical power counting. At a truly non-perturbative fixed point, the relevant operators are not known. Therefore, simple truncations that correctly model non-perturbative physics can be difficult to devise. It is in such setups that the concerted use of several techniques can be most useful; the IR regime of QCD constitutes an excellent example. Finally, at an interacting, but near-perturbative fixed point, canonical power counting constitutes a viable guiding principle to set up truncations. Here, near-perturbative refers to the fact that the spectrum of critical exponents exhibits deviations of $\mathcal O(1)$ from the canonical spectrum of scaling dimensions, but not significantly larger, in other words, the anomalous contribution to the scaling of operators is $\eta_{\mathcal{O}}\lesssim \mathcal{O}(1)$. 

The strategy that has (implicitly or explicitly) been followed for the choice of truncations for the Reuter fixed point has been based on the assumption of near-perturbativity. This motivates a choice of truncation based on canonical power counting. The self-consistency of this assumption has to be checked by the results within explicit truncations. Indeed, \cite{Falls:2013bv,Falls:2014tra,Falls:2017lst,Falls:2018ylp} find a near-canonical scaling spectrum in the $f(R)$ truncation. 
Moreover, \cite{Eichhorn:2018akn,Eichhorn:2018ydy,Eichhorn:2018nda} find close agreement of various ``avatars" of the Newton coupling, something that is not expected in a truly non-perturbative regime.

As a self-consistent truncation scheme appears to be available for quantum gravity, the apparent convergence of fixed-point results is a key goal. It is fair to say that the status of results is rather encouraging with regard to this question, see \cite{Lauscher:2001ya,Lauscher:2001rz,Reuter:2001ag,Lauscher:2002sq,Reuter:2002kd,Reuter:2003ca,Bonanno:2004sy,Reuter:2008wj,Reuter:2008qx,Daum:2008gr,Manrique:2008zw,Daum:2009dn,Manrique:2008zw,Daum:2009dn,Manrique:2010mq,Manrique:2010am,Harst:2011zx,Nink:2012vd,Becker:2014qya,Becker:2014pea,Nink:2015lmq,Pagani:2019vfm,Franchino-Vinas:2018gzr, Bosma:2019aiu, Knorr:2019atm}. This has given rise to the general expectation that the Reuter fixed point indeed exists in full theory space, and provides a universality class for quantum gravity. Nevertheless, it should be pointed out that due to the technically very challenging nature of these calculations, the inclusion of a complete set of curvature-cube operators remains an outstanding task. In the vertex expansion, higher order derivative terms are captured by momentum-dependent correlation functions, which exhibit robust evidence for the Reuter fixed point
\cite{Christiansen:2012rx, Christiansen:2014raa, Meibohm:2015twa, Christiansen:2015rva, Denz:2016qks, Christiansen:2017cxa, Christiansen:2017bsy, Eichhorn:2018akn, Eichhorn:2018ydy, Franchino-Vinas:2018gzr,Knorr:2018kog,Eichhorn:2018nda,Bosma:2019aiu}.

\subsection{Do backgrounds matter?}\label{sec:background}
\emph{...where we highlight the technical challenges one faces when attempting to reconcile the use of a local coarse-graining procedure with the background independence expected of a non-perturbative quantum gravity approach.}\newline

When setting up the Wetterich equation for gravity \cite{Reuter:1996cp} the background field formalism plays an essential role. The background metric $\gb_{\mu\nu}$ serves the double purpose of i) introducing a gauge fixing which is invariant under background-transformations, and  ii) introducing a regulator, as required to implement a local notion 
of coarse graining. At the same time, the decomposition of the physical metric into a fixed, but arbitrary background and fluctuations introduces a new symmetry, so-called split-symmetry transformations: the linear split \eqref{eq:LinearSplit} is invariant under
\begin{equation}\label{splitsym}
\gb_{\mu\nu} \mapsto \gb_{\mu\nu} + \epsilon_{\mu\nu} \, , \quad h_{\mu\nu} \mapsto h_{\mu\nu} - \epsilon_{\mu\nu} \, . 
\end{equation}
While actions of the form \eqref{eq:EffActgeo} are invariant under these transformations, the gauge-fixing and the regulator terms 
\begin{equation}\label{reg}
\Delta S_k = \int d^4x \sqrt{\gb} \, h_{\mu\nu}[R_k(-\bar{D}^2)]^{\mu\nu\kappa\lambda}h_{\kappa\lambda}\ ,
\end{equation}
with $\bar{D}_{\mu}$ denoting the covariant derivative constructed from $\gb_{\mu\nu}$ violate this symmetry. Thus $\Gamma_k[h_{\mu\nu}; \gb_{\mu\nu}] \equiv \Gamma_k[g_{\mu\nu}, \gb_{\mu\nu}]$ genuinely depends on two metric-type arguments.

Nevertheless, the gravitational effective average action \cite{Reuter:1996cp} provides a background-independent approach to quantum gravity. The background metric $\gb_{\mu\nu}$ is not an ``absolute element'' of the theory but rather a second, freely variable  metric-type argument which is determined from its own equations of motion. At the most conservative level this feature follows from standard properties of the background field method satisfied by the effective action $\Gamma$ and their extension to the effective average action $\Gamma_k$ \cite{Pawlowski:2003sk, Donkin:2012ud, Folkerts:2011jz, Morris:2016spn, Percacci:2016arh, Labus:2016lkh, Ohta:2017dsq, Nieto:2017ddk, Christiansen:2017bsy}. Alternatively, it has been proposed to achieve background independence not by quantization \emph{in the absence} of a background, but rather by quantization \emph{on all background simultaneously} \cite{Becker:2014qya}.
We now review these arguments.  

The fact that $\Gamma_k$ and the resulting effective action $\Gamma$ depend on two arguments allows to derive a background as well as a quantum equation of motion
\begin{equation}
\left. \frac{\delta \Gamma[h;\gb]}{\delta \gb_{\mu\nu}} \right|_{\gb = \gb^{\rm eom}, h=0} = 0 \, , \quad 
\left. \frac{\delta \Gamma[h;\gb]}{\delta h_{\mu\nu}} \right|_{\gb = \gb^{\rm eom}, h=0} = 0 \, . 
\end{equation}
The Ward identity following from the transformation \eqref{splitsym} then relates these two equations implying that a solution of one is also a solution of the other. This allows to fix $\gb$ in a dynamical way. In particular, it shows that at $k=0$ the background metric does not have the status of an absolute element. At finite values of $k$, the Ward identity satisfied by $\Gamma_k$ receives additional contributions from the regulator \eqref{reg} which introduce a genuine dependence on the background field. From these arguments, it is then clear that ``background independence'' is restored at $k=0$ only. 

The ``all backgrounds is no background'' proposal provides an extension of ``background independence'' to finite values of $k$.  The underlying idea is to describe (one single) background-independent quantum field theory of the metric through the (infinite) family of ``all possible'' 
background-dependent field theories that live on a non-dynamical classical spacetime. Each family member has its own classical metric $\gb_{\mu\nu}$ rigidly attached to the spacetime manifold.
For each given background $\gb_{\mu\nu}$, standard methods can be used to quantize the fluctuation fields $\Phi$. Repeating this procedure for all $\gb_{\mu\nu}$ yields
expectation values $\langle \cO \rangle_{\gb}$ which are manifestly $\gb$ dependent in general. 
Loosely speaking, the family of backgrounds, which is at the heart of background independence in the abstract sense of the word, should be regarded as  the set of all possible ground states, one of which will be picked dynamically.

Ultimately, the physical background metric that is present in the geometric phase of quantum gravity, is determined by the dynamics of the system  in a self-consistent fashion by solving the quantum equations of motion at finite $k$
\begin{equation}
\left. \frac{\delta \Gamma_k[h;\gb]}{\delta h_{\mu\nu}} \right|_{\gb = \gb^{\rm sc}_k, h=0} = 0 \, , 
\end{equation}
where the self-consistent background metric $( \gb^{\rm sc}_k)_{\mu\nu}$ is inserted. Hence, the expectation value of the metric is a prediction rather than an input. Notably, setting $(h=0, \gb = \gb^{\rm sc}_k)$ is a particular way of going ``on-shell'' (but not the only one). We refer to \cite{Christiansen:2017bsy, Reuter:2019byg} for further details.

Given these remarks, it is clear that future work must address the following challenges: \\

\noindent {\bf (1)} The different functional dependence of $\Gamma_k$ on $h_{\mu\nu}$ and $\gb_{\mu\nu}$ induces differences in the propagators for the fluctuation field and the background field. Thus,
the functional dependence of $\Gamma_k$ on $h_{\mu\nu}$ and $\gb_{\mu\nu}$ separately should be computed for a class of background metrics as broad as possible, as ultimately background independence can only be achieved if the dependence on the two distinct arguments of $\Gamma_k$ is disentangled cleanly.

For computational feasibility the existing calculations mainly employ either highly symmetric background geometries or the Seeley-DeWitt (early time) expansion of the heat-kernel which encapsulates only local (albeit universal) information \cite{Benedetti:2010nr}.  It is important to highlight that computations evaluating the left-hand side of the Wetterich equation at $h=0$ (i.e., equating fluctuation propagator and background propagator) can deform and/or remove fixed points and introduce unphysical zeros of beta functions \cite{Bridle:2013sra}.
\\
{\bf (2)} The difference between the $g_{\mu\nu}$ dependence of $\Gamma_k$ and its $\bar{g}_{\mu\nu}$ dependence, driven by the distinct dependence of regulator and gauge fixing on the two metrics, is encoded in the modified split Ward  or Nielsen identity resulting from \eqref{splitsym}. In principle, by solving the flow equation together with this Ward identity,
one would obtain a flow for a functional of a single metric.
In practice, the solutions of the Ward identity has only
been possible for the simplest approximations \cite{Donkin:2012ud,  Morris:2016spn, Percacci:2016arh, Labus:2016lkh, Ohta:2017dsq, Nieto:2017ddk}. 

\noindent {\bf (3)} When $\Gamma_k$ and with it $(\gb^{\rm sc}_k)_{\mu\nu}$ show a strong $k$ dependence, the effective spacetime is likely to possess multi-fractal properties which were argued to lead to a dimensional reduction in the ultraviolet \cite{Lauscher:2005qz,Reuter:2011ah,Rechenberger:2012pm,Calcagni:2013vsa} and to a ``fuzzy'' spacetime structure at even lower scales \cite{Reuter:2005bb,Reuter:2006zq,Pagani:2019vfm}. In the existing analyses the fractal-like properties were characterized in terms of ordinary, i.e., smooth classical metrics, the trick being that one and the same spacetime manifold was equipped not with one but rather the one-parameter family of classical metrics, $\{(\gb^{\rm sc}_k)_{\mu\nu}\}$.  As these fractal-like properties relate to the $k$ dependence of $\Gamma_k$, it is at present unclear whether an ``echo" of this behavior exists in the physical limit $k \rightarrow 0$. Investigating the full momentum dependence of $\Gamma_{k \rightarrow 0}$ can provide an answer to this question.
If there is, it should be a mostly negligible effect at scales relevant for current experiments.

In conclusion, the issue of the background dependence is a main obstacle to
progress in the application of the FRG to quantum gravity, both at the conceptual and technical level.

\section{Additional methods for asymptotic safety}
\label{sec:OtherApproaches}
\emph{... where we review other techniques used to search for asymptotic safety in gravity, including the $\epsilon$ expansion, numerical simulations, tensor models,
and stress the benefits of using multiple methods.}\\

Sections \ref{sec:convergenttruncations} and \ref{sec:background} have highlighted the technical challenges one faces when employing the FRG to study asymptotically safe gravity. Therefore, there is a strong case for the use of complementary methods,
especially those where background independence can be implemented, such as Regge calculus or random lattice techniques, as well as specific tensor models. 
Due to the rather different nature of the systematic errors in these approaches, this simultaneously addresses the challenge linked to the convergence of truncations. 
Furthermore, other techniques may be better suited to explore the complete phase diagram of quantum gravity potentially including pre-geometric phases. 

Historically, the starting point for studies of asymptotically safe gravity has been the $\epsilon$ expansion around $d=2$, 
 \cite{Gastmans:1977ad,Christensen:1978sc,Kawai:1989yh,Kawai:1992np,Kawai:1993mb}, which has been pushed to two-loop order in \cite{Aida:1996zn}. 
It has been shown that the Reuter fixed point in $d=4$ dimensions is continuously connected to the perturbative fixed point seen in $2+\epsilon$ spacetime dimensions \cite{Reuter:1996cp,Reuter:2001ag}.
The connection between Asymptotic Safety and Liouville gravity in $d=2$ dimensions has been made in \cite{Nink:2015lmq}.
An (off-shell) gauge and parameterization dependence, as exhibited by truncated FRG studies, is also present in the $\epsilon$ expansion. Higher-loop terms are required in order to resum the $\epsilon$ expansion for the critical exponent to learn about the $d=4$-dimensional case. This appears to be merely a technical challenge, to which the advanced techniques developed in the context of supergravity \cite{Bern:2008qj}  might potentially be adapted.\\
In line with the near-perturbative nature of the fixed point in $d=4$, expected from FRG studies \cite{Falls:2013bv,Falls:2014tra,Falls:2017lst,Falls:2018ylp,Eichhorn:2018ydy}, a Pad\'e resummation might yield a fixed point that is continuously connected to the fixed point in the vicinity of two dimensions. 

Lattice approaches provide access to a statistical theory of random spatial geometries, thereby being in a position to provide evidence for or against asymptotic safety in the Euclidean regime. 
There are two main ways in which discrete random geometries are explored: One can hold a triangulation fixed and vary the edge lengths, as in Regge calculus, or hold the edge lengths fixed but vary the triangulation, as in dynamical triangulations.
The latter have developed in two research branches:  
Euclidean Dynamical Triangulations \cite{Ambjorn:1996ny}, and Causal Dynamical Triangulations \cite{Ambjorn:2012jv,Loll:2019rdj}. 
\\
Regge calculus (see \cite{Hamber:2009mt} for a review) based on the Einstein-Hilbert action is subject to the well-known conformal factor instability, which requires an extrapolation in order to extract information about a critical point, see the discussion in \cite{Hamber:2015jja}. With this caveat in mind, indications for asymptotic safety are found in Monte Carlo simulations of Regge gravity \cite{Hamber:2015jja}  based on the Einstein-Hilbert action. Testing the effect of additional, e.g., curvature-squared operators, which could correspond to additional relevant directions and have an important impact on the phase structure, is an outstanding challenge in Regge gravity.  A first comparison of scaling exponents obtained with the FRG to the leading-order exponent in Regge gravity can be found in \cite{Falls:2014zba, Falls:2015cta,Biemans:2016rvp}.
\\
In the case of Causal Dynamical Triangulations (CDT), the configuration space includes only configurations that admit a Wick rotation, see \cite{Loll:2019rdj} for a review. Therefore, an analytical continuation to a Lorentzian path integral is in principle possible. In two dimensions, one can solve CDT analytically. Owed to the fact that in this case there are no local degrees of freedom, it has been shown in \cite{Ambjorn:1998xu} and \cite{Ambjorn:2013joa,Glaser:2016smx} that the Hamiltonian appearing in the continuum limit agrees with the one for two-dimensional continuum quantum gravity  and Horava-Lifshitz gravity \cite{Horava:2009uw}, respectively. Moreover, Liouville gravity can be recovered by allowing for topology change of the spatial slices \cite{Ambjorn:1998xu}. It has been stressed in \cite{Loll:2019rdj} though that the equivalence of CDT and Horava-Lifshitz gravity may not extend beyond the two-dimensional case.

In higher dimensions, one searches for the continuum limit numerically.
In practice, evidence for several \cite{Ambjorn:2011cg,Ambjorn:2012ij} second-order phase transition lines/points exists in numerical simulations, both in spherical and toroidal spatial topology.
The large-scale spatial topology does not appear to impact the phase structure \cite{Ambjorn:2018qbf}, but can actually improve the numerical efficiency of the studies, as observed in \cite{Ambjorn:2019lrm}. 
The higher-order transition can be approached from a phase in which several geometric indicators (spatial volume of the geometry as a function of time \cite{Ambjorn:2004qm}; Hausdorff dimension and spectral dimension \cite{Ambjorn:2005db}) signal the emergence of a spacetime with semi-classical geometric properties. The properties of the continuum limit remain to be established, as the process of following RG trajectories along lines of constant physics towards the phase transition has not yet led to conclusive results regarding asymptotic safety \cite{Ambjorn:2014gsa,Ambjorn:2016cpa}. 
\\
In Euclidean Dynamical Triangulations (EDTs), the configuration space differs from CDTs, as configurations do not in general admit a Wick rotation. This gives rise to spatial topology change and the proliferation of so-called ``baby universes". Early work \cite{Ambjorn:1991pq,Agishtein:1991cv,Catterall:1994pg,Bilke:1996ek} has not shown a higher-order phase transition \cite{Bialas:1996wu,deBakker:1996zx,Catterall:1996gj}. The inclusion of a measure-term has led to the hypothesis that the first-order transition line could feature a second-order endpoint, and some evidence exists that
the volume profile of the ``emergent universe" approaches that of Euclidean de Sitter, i.e., a sphere, as one tunes towards the tentative critical point \cite{Laiho:2016nlp,Ambjorn:2013eha}. This measure term could be reinterpreted as a sum of higher-order curvature invariants \cite{Laiho:2016nlp,Ambjorn:2013eha} contributing to the action.
The investigation \cite{Ambjorn:2013eha} was unable to corroborate the appearance of a second-order endpoint though.
\\
Finally, dynamical triangulations can be encoded in a purely combinatorial, ``pre-geometric" class of models, so-called tensor models \cite{Ambjorn:1990ge,Godfrey:1990dt,Sasakura:1990fs, Gurau:2009tw,Gurau:2011aq,Gurau:2011xq, Bonzom:2012hw,Carrozza:2015adg}, 
that attempt to generalize matrix models \cite{Ginsparg:1993is} for 
two-dimensional gravity to the higher-dimensional case.
FRG tools which interpret the tensor size $N$ as an appropriate notion of ``pre-geometric" (i.e., background-independent) coarse-graining scale \cite{Eichhorn:2013isa}, allow to recover the well-known continuum limit in two-dimensional quantum gravity within systematic uncertainties related to truncations \cite{Eichhorn:2014xaa}. First tentative hints for universal critical behavior in models with 3- and 4-dimensional building blocks have been found \cite{Eichhorn:2018phj,Eichhorn:2019hsa}. This method could in the future provide further evidence for asymptotic safety, see \cite{Pereira:2019dbn} for a discussion, once the systematic uncertainties are reduced by suitable extensions of the truncation, and an understanding of the emergent geometries has been developed.\\
More broadly, the framework of the Renormalization Group and the notion of a universal continuum limit linked to a fixed point have recently been gaining traction in several approaches to quantum
gravity, including group field theories \cite{Oriti:2006se,Rivasseau:2011hm,Carrozza:2016vsq} as well as spin foam models \cite{Dittrich:2014ala}. Accordingly, the concept of asymptotic safety might play an important role in several distinct approaches to quantum gravity. In particular, in spin foams, a search for interacting fixed points in numerical simulations has  started recently in reduced configuration spaces, see, e.g., \cite{Delcamp:2016dqo,Bahr:2016hwc,Bahr:2018gwf}.

In summary, the further development and application of a broad range of tools to explore asymptotic safety could be key to gain quantitative control over a potential fixed point, establish its existence and to develop robust links to phenomenology, which rely on a good understanding and control over systematic errors within various techniques.

\section{Running couplings}
\label{sec:runningcouplings}

\subsection{A clarification of semantics}\label{sec:semantics}

\emph{...where we clarify that the term ``running coupling" is used
with different meanings in different contexts.}
\\

Much of the current work on asymptotic safety of gravity uses techniques and jargon that are more common in statistical than in particle physics. This concerns even basic notions such as the RG. 
If one aims at detecting asymptotic safety by means of standard perturbative particle physics
observables, there is thus much room for misunderstanding.

The RG was used in particle physics largely as a tool to resum ``large logarithms'', terms in the loop corrections 
to physical observables of the form $\log(p/\mu)=\log(p/\Lambda) +\log(\Lambda/\mu)$, where $p$ is a momentum, $\mu$ a reference scale and $\Lambda$ a UV cutoff.
From the way they emerge, the beta functions that resum the large logs are just the coefficients of the logarithmic divergences $\log(\Lambda/\mu)$.
One important feature of these logarithmic terms is that their coefficients are ``universal'', up to next-to-leading non trivial order (NLO) in the coupling expansion. 
This entails two things: on the one hand, it means that, up to NLO, they are independent of the way one computes them 
\footnote{They are almost always derived in dimensional
regularization, which for technical reasons is the most convenient method, e.g. it respects gauge symmetries.}.
On the other hand, one can use them to ``RG improve'' {\it any} tree level observable, and one is guaranteed to obtain the correct result
(not the full result, of course, but the part that comes from
calculating and then resumming the logs).
Here by ``RG improvement'' we mean the substitution of the running coupling into a tree-level expression,
and the subsequent identification of the RG scale with an appropriate physical scale of the system.
\footnote{
For example in a process $e^+e^-\to e^+e^-$ at center of mass energy $\sqrt{s}>>m_e$ at $n$-loop order the renormalized leading contribution with subtraction scale $\mu$ is proportional to the tree level cross-section (at scale $\mu$) times $\sum_{l\le n} [\alpha(\mu) c\, \log \frac{s}{\mu^2}]^l$, which shows that the most convenient choice is $\mu=\sqrt{s}$.}
If one demands these properties of a running coupling, then
one would say that {\it only dimensionless couplings can run}.
Dimensionful couplings have power divergences that are simply
subtracted in perturbation theory. 
In line with these arguments, it has been pointed out in \cite{Anber:2011ut,Donoghue:2019clr}, that the one-loop corrections to gravity-mediated scattering amplitudes cannot be obtained from applying the RG improvement to  Newton's coupling.

In Wilson's non-perturbative approach to renormalization, all possible terms consistent with symmetries are present in the action. Quite often, the Wilsonian momentum cutoff has a direct physics interpretation, e.g., as lattice spacing in condensed-matter applications (with a relation to the Kadanoff block-spinning \cite{Kadanoff:1966wm} underlying Wilson's renormalization idea), and as the mass of states that are ``integrated out'' in effective field theories. In lattice gauge theories the Wilsonian momentum cutoff is finally removed (in the continuum limit), but keeps its physics interpretation similar to the condensed-matter applications at intermediate stages. Nevertheless, the momentum cutoff is treated mathematically as an independent variable, and all couplings in the Wilsonian action depend on it.
Apart from a few relevant parameters to be tuned to criticality, the remaining set of ``running couplings''  is not constrained by the demands of universality; still, this notion of running couplings remains  also valid at the non-perturbative level.

The relation between the two definitions of the RG is this:
At energy scales much higher than all the masses, 
the leading- and next to leading-order terms of the
perturbative beta functions, that are independent of the
renormalization scheme,
can also be obtained from the Wilsonian RG and are
independent of details of the coarse-graining scheme.
In particular, the one-loop terms can be easily found
from (\ref{eq:tracelog}). 
The recovery of 2-loop terms from the FRG has been addressed, e.g., in \cite{Papenbrock:1994kf, Reuter:1997gx, Pernici:1998tp, Kopietz:2000bh, Pawlowski:2001df, Gies:2002af, Morris:2005tv}. 
At energies comparable to the masses, the beta functions extracted from the Wilsonian RG include threshold effects which encode the automatic decoupling of massive modes from the flow at scales below the mass. This is an advantage over setups in which this decoupling is not accounted for automatically and must instead be done by hand.

If one accepts the more general Wilsonian definition of running coupling, then the statement ``dimensionful couplings cannot run'' translates into the statement that the Wilsonian running of dimensionful couplings does not carry the same direct physical meaning as the running of dimensionless ones. Nevertheless, to encode physics correctly within a Wilsonian setup, the running of dimensionful couplings is critical and cannot be neglected, since in a massive scheme operators mix non trivially.

To be more specific, one can consider what happens at second order phase transitions. The generic power-like running of the Wilsonian couplings in the FRG approach is in general non-perturbative and its calculation is limited only by the approximations.
At the fixed point, the couplings have non-universal values (depending on the details of the microscopic theory), but there are also universal quantities which can be extracted from the flow close to criticality.
These are the same for very different physical systems belonging to the same universality class. 
The power-like divergences are associated to non-universal features such as the position of the fixed point and of the critical surface (see, e.g.~\cite{Wetterich:2019qzx}).  
For example, the power quadratic divergence in systems belonging to the Ising universality class is related to the critical temperature $T_c$, which varies from one material to another. If one is interested in this physical information, the accurate scaling of the corresponding quadratic composite operator or the behavior of the two-point function should be determined.

Similar considerations may apply in quantum gravity, where the running Planck mass (the coefficient of the ``$R$'' operator in the effective Lagrangian) is a non-universal quantity which is just one of the parameters defining the position of a possible UV fixed point and of the critical surface containing it. 
Note that in an asymptotically safe theory of quantum gravity, 
the physics is related not just to the UV fixed point, but to the particular renormalized trajectory flowing away from it towards lower energy scales. Therefore it depends indirectly on all such Wilsonian (dimensionful) couplings. Observables, as already discussed, are computed at $k=0$ on the on-shell configurations and are mostly sensible to a number of non-universal parameters related to the finite number of relevant directions, including the (flowing) Planck mass.
We shall discuss in Sect.~\ref{sec:CorrCoupOb} how one could define the effective couplings.

\subsection{Remarks on dimensional regularization}
\emph{...where we explain in which cases some care is required for the correct interpretation of results achieved within dimensional regularization.}\\

A seemingly technical point where the Wilsonian RG approach differs from a
perturbative particle-physics perspective is the
regularization of quantum modes. While the FRG works with explicit
momentum-space regulators (or spectral regulators of curved-spacetime
Laplacians), conventional perturbation theory mostly uses dimensional
regularization for reasons of convenience. Physics must
not depend on the choice of the regularization scheme, 
hence it is an obvious question as to whether dimensional regularization can also be brought to work
in a FRG context and for the asymptotic-safety scenario of gravity.

In fact, one-loop results for power-counting marginal operators quadratic in
the curvature with dimensionless couplings exhibit the expected universality
\cite{Codello:2006in,Avramidi:1985ki,deBerredoPeixoto:2004if}. However, this
is no longer true for the RG running of power-counting relevant and irrelevant
operators, simply because they do not feature the same degree of
universality. Even worse, dimensional regularization is blind to power
divergencies and hence acts as a projection onto logarithmic divergences
appearing as $1/\epsilon$ poles.
For such reasons, Weinberg calls dimensional
regularization ``a bit misleading'' in the context of asymptotically safe theories \cite{Weinberg:1980gg}.

Dimensional regularization relies on the virtues of analytic continuation. Hence, its application requires to pay attention to the analytic structure of a problem at hand. This is well known, for instance, from  non-relativistic scattering problems where a naive application of dimensional regularization fails because of a different analytic structure of the propagators and more care is needed to apply analytic continuation methods to regularize and compute observables \cite{Phillips:1997xu,Beane:1997pk}. The same is true for computations in large background fields where a naive straightforward application of dimensional regularization is
not possible, but requires a careful definition in terms of a dimensionally
continued propertime or $\zeta$ function regularization
\cite{Brown:1976wc,Luscher:1982wf}. The latter techniques can be linked to heat-kernel methods and allow to  access information related to power divergences \cite{Gies:2014xha}.

As most computations for asymptotically
safe gravity are performed in ``large backgrounds'', i.e., in a fiducial
background spacetime, a proper use of dimensional regularization would similarly require a definition in terms of, e.g., a propertime or $\zeta$ function definition based on the heat kernel. In fact, approximations of the FRG have been mapped onto a propertime representation (propertime RG). Applications to gravity do lend further support to the existence of the Reuter fixed point and the asymptotic-safety scenario \cite{Bonanno:2004sy}.

 \subsection{Correlation functions and form factors}\label{sec:CorrCoupOb}

\emph{...where we clarify the distinction between RG scale dependence and physical scale dependence within the FRG context. We further detail how the physics of asymptotically safe theories is encoded in momentum-dependent correlation functions and form factors, discuss the definition of non-perturbative running couplings and the construction of observables from these objects.}\\

The idea of the Wilsonian renormalization group is to solve the theory by integrating out quantum fluctuations, 
one (covariant) momentum shell at the time. 
It is crucial to distinguish the $k$-dependence from the dependence on physical scales.
In the FRG approach governed by the Wetterich equation \eqref{FRGE} with an infrared cutoff, 
general correlation functions 
\begin{align}\label{eq:CorrFun}
{\cal C}_k(p_1,...,p_n)= \langle \Phi_{i_1}(p_1) \cdots \Phi_{i_n}(p_n)\rangle_k
\end{align}
are trivial for all (covariant) momentum scales $p_i^2/k^2 \ll 1$, and carry the momentum dependence of the 
full theory for all momentum scales $p_i^2/k^2 \gg 1$ and large scattering angles. 
Here, trivial means that for $p_i^2/k^2 \ll 1$ the correlation functions are that of a theory with a mass gap $m_\textrm{gap}^2\gtrsim k^2$: the quantum dynamics dies off with powers of $p_i^2/k^2$. In asymptotically safe theories, these correlation functions will exhibit indications of quantum scale invariance at large $p_i^2/k^2$. 

Evidently, the correlation function ${\cal C}_k(p_1,...,p_n)$ is a highly non-trivial function of all $p_i$, other physics scales in the theory, such as mass scales, and the cutoff scale $k$. The latter is instrumental for the transition from the full quantum dynamics of the theory to the trivial one in the gapped regime. This non-trivial behavior is complicated by the 
fact that the $n$-point correlation functions carry $n$ momenta $p_i$ with $i=1,...,n$. This results in a multiscale problem, 
unless we restrict ourselves
to a symmetric point with $p_i^2=p^2$. 
We also remark that both UV and IR regimes may exhibit asymptotic power-law momentum scaling or anomalous scaling and the momentum and cutoff dependence in the transition regime at $p_i^2/k^2\approx 1$ is in general highly non-trivial. In particular, the momentum dependence is typically more general than a logarithmic one. 

In Sect.~\ref{sec:Observables} we introduce diffeomorphism invariant observables as spacetime integrals over correlators akin to the one in Eq.~\eqref{eq:CorrFun}, cf.~Eq.~\eqref{eq:CurvatureExp}, or $S$ matrix elements via the proper background vertices $\bar\Gamma_k^{(\bar\Phi_{i_1}\cdots \bar \Phi_{i_n})}$. To compute such observables, in the FRG approach to asymptotically safe gravity we first have to compute the proper vertices of the fluctuation fields, $\Gamma_k^{(\Phi_{i_1}\cdots \Phi_{i_n})}$. 
Their scale- and (covariant) momentum dependence 
indirectly encode the physics of asymptotically safe gravity despite not being observables themselves. An important step towards observables is made by considering running couplings, that are renormalization group invariant combinations of the \textit{form factors} or \textit{dressings} of these vertices as defined in standard gauge theories and scalar and fermionic QFTs. These are defined from the $k$- and momentum-dependent vertices  together with appropriate factors of the wave-function renormalizations.
For instance, in the case of scalar and fermionic QFTs, these are directly related to $S$ matrix elements.  
In turn, in gauge theories such as QCD they lack gauge invariance but nonetheless carry important physics information: In QCD these running couplings derived from the proper vertices of the fluctuating  or background fields give direct access to the momentum scaling in the perturbative regime as measured by 
high-energy experiments, see, e.g., \cite{Tanabashi:2018oca}. Further, non-perturbative physics,  such as the emergence of the confinement mass gap, is also captured by these running couplings, see, e.g., \cite{Mitter:2014wpa, Cyrol:2016tym, Cyrol:2017ewj}. 

This implies that the momentum dependence of these running couplings at $k=0$ provides rather non-trivial physics information. In asymptotically safe gravity, it can in particular be used to identify scaling regimes in the UV and the IR as well as the transition scale: In \cite{Christiansen:2012rx, Christiansen:2014raa, Christiansen:2015rva, Denz:2016qks} non-perturbative generalizations of the Newton coupling $G_k^{(n)}(p^2)$ with $n = 3,4$, defined from the $n$-point functions, have been computed from combinations of the proper two-, three- and four-point functions of the fluctuation fields in a 
flat background and all cutoff scales. For the generalization to the case with matter, see \cite{Christiansen:2017cxa,Eichhorn:2018akn,Eichhorn:2018ydy,Eichhorn:2018nda}. In these calculations, the dependence on the $n-1$ momenta of an $n$-point vertex has been simplified by going to the momentum-symmetric point, allowing the definition of a running coupling that depends on a single momentum.
A flat background, as used in the above studies is of course a first step towards a comprehensive understanding of the physical scale dependence of quantum gravity.
For first steps towards an extension to generic background see \cite{Christiansen:2017bsy,Burger:2019upn}. 

On a generic background, the dependence on physical scales can also be captured in the language of form factors. In the background effective action 
these form factors  appear naturally, see \cite{Barvinsky:1990up},  
\begin{align}\nonumber 
\bar \Gamma[g_{\mu\nu}] = & \int d^4x \sqrt{g} \big[ f(R)  + f_1(R_{\mu\nu} R_{\mu\nu}) \\[1ex] 
& \; + C_{\mu\nu\rho\sigma}\, W^{\rm T}(\Delta)\, C^{\mu\nu\rho\sigma} - R\, W^{\rm R}(\Delta)\, R + \cdots \big] \, , \label{Geff}\end{align}
Equation \eqref{Geff} also summarizes concisely the approximation considered so far for the background effective action. The corresponding form factors $W^{\rm R}$ and $W^{\rm T}$ have been computed in \cite{Knorr:2019atm}. 
Note that \eqref{Geff} can also be understood as the dynamical effective action in the diffeomorphism invariant 
single metric approach put forward in \cite{Wetterich:2016qee, Wetterich:2017aoy, Pawlowski:2018ixd, Wetterich:2019qzx, Wetterich:2019zdo}. There it has been argued that the physical gauge there facilitates the direct physics interpretation of form factor such as $W^{\rm T}$ and $W^{\rm R}$.  

Both within the language of momentum-dependent correlation functions as well as with form factors,  the asymptotically safe regime, the transition regime and a long infrared regime with classical scaling have been identified. The results are rather promising and open a path towards the computation of observables or their local integral kernels. Still, the approximations used so far do in particular not sustain large curvatures and have to be upgraded significantly.

\section{Observables}
\label{sec:Observables}
\emph{...where we  emphasize  the necessity to investigate observables in order to make quantum gravity testable,
and discuss three possible classes of observables.}\\

The physical behavior of a system is probed through observables. 
While their definition and construction is not a problem in many interesting cases of quantum and statistical field theories in flat, and possibly some specific classical curved spacetimes,
it is in general very difficult to define meaningful observables in quantum gravity.
To begin with, already in classical gravity diffeomorphism invariance
makes the notion of a spacetime point unphysical and hence
implies that there cannot exist any local observable:
any gauge invariant observable must be the 
integral of a scalar density over all spacetime.
The situation is somewhat better in the presence of matter,
for example it makes sense to define the value of the scalar
curvature at the position of a particle,
or at a point where certain matter fields have predetermined values
\cite{Westman:2007yx}. 
These observables are however difficult to work with in practice.
These problems persist in quantum gravity,
see, e.g., \cite{Brunetti:2013maa}. Nevertheless the construction of observables remains a crucial task.

In the following, we will focus on observables in the sense of quantities that are of direct phenomenological relevance. These often rely on introducing a (dynamically generated) background that provides a suitable notion of locality.
The type of observables that one will consider depends very strongly
on the type of observations that one has in mind.
We will distinguish three possible classes of observations
that could be used to test asymptotic safety.

\subsection{Particle physics at the Planck scale}\label{sec:CrossPart}

The first is appropriate when we imagine living 
in a macroscopic classical spacetime and probing 
its short distance structure by some ``microscope''
of the kind that is used in particle physics.
For example, we could try to directly measure
scattering cross-sections and decay rates
at Planckian scale or beyond.
In this case the issue of diffeomorphism invariance is
circumvented by postulating the existence of an 
asymptotically flat background, which is necessary
in order to define the appropriate notions of
particles and asymptotic states. The validity of this postulate remains to be investigated in a given quantum theory of gravity.
In principle, the integral kernels of these particle-physics observables can be constructed from the proper vertices of the background effective action $\bar \Gamma_k^{(\bar\Phi_{i_1}\cdots \bar\Phi_{i_n})}$ for an asymptotically flat spacetime, see \cite{Abbott:1983zw}. We provide some details on the calculation of these quantities in Sec.~\ref{sec:CorrCoupOb}.
Indeed, the original formulation of Asymptotic Safety by Weinberg was formulated in these terms: 
as stated in \cite{Weinberg:1976xy,Weinberg:1980gg}, ideally,
the couplings whose running one wants to study
should be defined directly in terms of such observables.
However, most of the actual work on Asymptotic Safety is based on
the running of parameters in the Lagrangian,
that are not directly observable
or not even directly related to observables.
Assuming that this notion makes sense, 
measurement of the $S$ matrix at the Planck scale and beyond would give the most direct and unambiguous test of Asymptotic Safety.
Unfortunately, neither the theoretical nor the experimental
sides of the comparison are available.
In settings with extra dimensions, scattering cross sections have been calculated within the framework of RG improvement 
\cite{Litim:2007iu,Gerwick:2011jw,Dobrich:2012nv}, see Sec.~\ref{RGimpro} for a discussion of the potential pitfalls of this procedure.
With current technology,
these observables are also unlikely to ever be measured.
Furthermore, the postulate of an asymptotically flat background
leaves out many situations that are of interest
in the context of quantum gravity.

\subsection{Low-energy imprints}\label{sec:LowIm}

A second possibility,
still closely related to the world of particle physics,
but not requiring Planckian energy,
is the observation of properties of the low-energy world
that could carry an imprint of asymptotically safe quantum gravity.
One can distinguish two sub-cases, that we shall
refer to as  ``higher-order observables" and ``marginal observables".
Both sets of observables are most directly calculable
if one assumes a ``great desert''
between the Planck and the Fermi scale.
Else, one requires a specific model for the intervening physics.\\[-2ex]

\noindent \emph{i)} The high energy theory will leave
traces in the low energy effective field theory in the form
of higher order operators that are suppressed by inverse powers
of the high scale. In particular, higher-order matter self-interactions are very likely both nonvanishing and irrelevant in the UV, 
if an asymptotically safe matter-gravity fixed point exists \cite{Vacca:2010mj,Eichhorn:2011pc,Eichhorn:2012va,Eichhorn:2016esv,Eichhorn:2017eht,Christiansen:2017gtg}. 
This results in predictions for these higher-order couplings in the IR.
The separation of scales between the Planck scale and IR scales
is so large that, typically, these quantum-gravity effects 
are unmeasurably tiny.
Still, one may hope that there exists a signature that is
forbidden in any non-gravitational process and that becomes
detectable under  rather unexpectedly favorable circumstances. \\[-2ex]

\noindent \emph{ii)} The other, significantly more promising, possibility is that some
gross features of the low energy world, probed at present or future colliders  and linked to canonically marginal couplings, i.e., 
dimensionless operators,  could be directly ``explained'' by properties of a UV-complete quantum theory of gravity and matter. 
This is due to the fact, explained in section~\ref{FRG},
that Asymptotic Safety may yield more predictions than a perturbatively renormalizable model.
In the gravitational sector, this mechanism may not lead to testable predictions: here only a handful of parameters are experimentally accessible and there are essentially no constraints on the value of the curvature-squared couplings. In the matter sector, this picture changes completely. In this case literally thousands
of observables
are available, depending on at least two dozen free parameters. 
Some of these canonically marginal couplings could become irrelevant directions of an asymptotically safe gravity-matter model.
First tentative hints have been obtained
in this direction, for example a proposed scenario for a prediction of the Higgs mass
\cite{Shaposhnikov:2009pv} and a calculation of the top mass \cite{Eichhorn:2017ylw}, the Abelian gauge coupling \cite{Harst:2011zx,Eichhorn:2017lry} and the bottom mass \cite{Eichhorn:2018whv}.  These are obtained in comparatively small truncations and are subject to the assumption that Euclidean results carry over to Lorentzian gravity-matter systems.
These are of course not ``smoking guns'' for Asymptotic Safety,  but it is not unreasonable to expect that a microscopic description of quantum gravity constrains the features of a matter sector that can consistently be coupled to it. 
Indeed, the swampland program in string theory is based upon the same assumption. 
Ultimately, one could hope to arrive at an extended list of calculable properties of matter models from various quantum gravity theories, allowing to rule out some of the latter observationally without the need to probe Planck-scale physics directly.
It  therefore seems worthwhile to  more systematically develop
the predictions that an asymptotically safe theory of gravity and matter
can make for low-energy observables.  In particular, the dark-matter sector could allow to make genuine predictions \cite{Eichhorn:2017als,Reichert:2019car,Hamada:2020vnf}, in contrast to the consistency tests that the already measured properties of the Standard Model provide.
We shall discuss in Sec.~\ref{link_eft_as} how such effects could be calculated.

\subsection{Asymptotically safe cosmology}\label{sec:ASCos}

The third class of observations is related to cosmology.
As long as a Friedmann-Robertson-Walker (or some other)
background is a good approximation,
there is a well-developed machinery for the
treatment of fluctuation correlators
\cite{Brunetti:2016hgw}.
At the formal level, observables in quantum gravity are given by
integrated correlators, for example spacetime integrals of $n$-point correlations of the Ricci scalar 
\begin{align} \label{eq:CurvatureExp}
{\cal O}_R^{(n)}= \langle \prod_i \int d^4 x_i \sqrt{g(x_i)}\, R(x_1) \cdots R(x_n)\rangle\,.  
\end{align} 
Naturally, while the spacetime dependent curvature fluctuations in \eqref{eq:CurvatureExp} are not observables  themselves, they carry the physics information encoded in its spacetime or momentum-scale dependence.

Inflation is believed to occur at sub-Planckian energies,
but it may be close enough to a fixed-point regime to be directly 
influenced by it.  Further, in settings like Starobinsky inflation, higher-order operators in the gravitational theory actually drive inflation. Along this line, it was explored in \cite{Gubitosi:2018gsl} whether the freedom in the $R^2$ coupling offered by Asymptotic Safety can be used to realize Starobinsky inflation giving power spectra compatible with present observations. Moreover, there are some tentative hints that quantum-gravity effects typically drive scalar potentials towards flatness, see, e.g., \cite{Narain:2009fy,Eichhorn:2017als,Pawlowski:2018ixd}, and generally impose strong constraints on the inflationary potential 
that is usually introduced in a rather ad-hoc manner, see also~\cite{Tronconi:2017wps}.
In a more unorthodox approach to early cosmology, the idea is being explored that quantum gravity directly solves the horizon, flatness and monopole problems and generates the appropriate spectrum of fluctuations without the need for additional degrees of freedom together with an ad-hoc potential. In particular, in \cite{Lehners:2019ibe} it has been demonstrated that an action including all gravitational four-derivative invariants leads to the suppression of spacetime configurations with an initial singularity as well as anisotropies and inhomogeneities.
In the early universe the usual flat space QFT machinery is not available and
one has to use different observables that are geared to
high temperature/high curvature situations.
Then one may hope that features of the fixed point such as
scaling exponents and OPE coefficients - that in statistical physics are generally considered measurable 
physical quantities - could leave an imprint in these cosmological observables.

\subsection{Remarks}\label{sec:Wrapup}

As with other situations where non-perturbative physics is involved,
one could try to cross-check results obtained with continuum QFT
methods with lattice studies.
It is worth mentioning that also in lattice 
approaches to quantum gravity, 
observables are very hard to define
and especially to implement in the simulations, see, e.g., \cite{Klitgaard:2018snm} for encouraging recent results.
This is in stark contrast to the large number of
observables that can be defined in the presence of
an asymptotically flat background.

Finally, let us recall that in other approaches to quantum gravity
such as LQG, ``geometrical'' observables
such as lengths, areas, volumes, and curvatures have played an important role.
These have also been discussed to some extent in Asymptotic Safety, \cite{Becker:2019tlf}, and can be computed with a flow equation for composite operators \cite{Pawlowski:2005xe, Igarashi:2009tj, Pagani:2016pad, Pagani:2016dof, Becker:2019fhi, Houthoff:2020zqy, Kurov:2020csd}.
While presently it is not clear what type of measurement
is required to access such observables, they can be used to explore whether different approaches to quantum gravity give rise to universal physical results.  Further, such geometrical observables have been used in \cite{Falls:2017cze} to set up a physical renormalization scheme.

\section{Relation of asymptotic safety to the effective-field theory approach}
\label{sec:AS-EFT}

\subsection{Asymptotic safety and Effective Field Theory}\label{link_eft_as}
\emph{...where we discuss the relation of the EFT framework to Asymptotic Safety and also outline a strategy how to devise approximations in which the link between the two descriptions can be established in practice.}\\

The framework of EFT is pervasive in modern particle physics. 
EFT is based on an expansion in $E/M$, where $E$ is the typical
energy scale of the experiment, and $M$ is the scale above which the EFT
  description may no longer be meaningful.
In EFT one finds that higher loop corrections are suppressed
by higher powers of $E/M$, so that the tree level and one loop
are usually enough to explain most of the phenomenology, provided the system is indeed perturbative in nature, as happens to be the case in many particle-physics applications.
Physical predictions are possible even when the theory is not
perturbatively renormalizable, as long as one considers only low-energy observables and assumes that the dimensionless counterparts of all couplings are roughly of $\mathcal{O}(1)$.

Einstein gravity is a paradigmatic example
of this point of view.
It is perturbatively non-renormalizable, but one can still reliably compute
observables in perturbation theory, as long as they are not affected by the
higher-derivative terms in the action, whose coefficients
are not calculable.
This is the case for some non-analytic parts of scattering
amplitudes.
The calculation of the quantum corrections to the Newtonian potential
is the most reliable calculation ever performed in quantum gravity \cite{Donoghue:1993eb}.
It is also the most accurate, since the separation of scales 
between the characteristic
scale of the theory (the Planck scale) and the scale where
one performs experiments (even at the LHC) is the largest of any EFT,
so that loop corrections are suppressed by enormous factors.
In this way, every test of Einstein's GR is also a 
test of this EFT of gravity.

In view of this, the motivation for asymptotic safety is twofold:
first, to have predictions for what happens at and beyond
the Planck scale and second, the promise of increased predictivity, in particular also at lower energies.
This is especially desirable in the presence of standard-model or
beyond-standard-model matter which is or might be detected in present and
future colliders or, e.g., in dark-matter detection experiments.
We stress again that  the enhanced predictivity comes from the fact that asymptotic safety selects a class of RG trajectories
which are expected to be parametrized by only a few free parameters.
In principle, all the remaining coefficients in the effective action are calculable,
including the coefficients of local higher dimensional operators that
appear perturbatively divergent and are therefore not calculable in the EFT.

When one follows a realistic RG trajectory from the UV fixed point, crossing the Planck scale and moving towards the IR, 
one must eventually arrive in the immediate neighborhood of the
free-theory fixed point of Einstein theory, which is the domain
where EFT is applicable.
In this regime, all the predictions of EFT must still hold true.
Indeed, in the FRG formalism, the loop expansion can be reconstructed
systematically by expanding the right-hand side of the equation in powers of
$\hbar$, cf.~Eq.~(\ref{eq:tracelog}). This is usually not done, because there are already
other methods that are perhaps better suited for this task;
but in principle, the FRG can reproduce all the results of the EFT in this way.

In practice, constructing a flow that links the description of the fixed point, which might or might not be near-perturbative, to the perturbative low-energy regime after potentially passing through a more strongly-coupled transition regime, is a challenge.
A possible strategy to deal with this complex problem is to figure out which parts of the flow can be captured by perturbation theory, and then use different tools (perturbation theory, one's favourite FRG approach) in the respective regime
so as to obtain maximally reliable predictions of the observables.  In order to link the description in terms of the FRG for the effective average action $\Gamma_k$ to the perturbative EFT setup, one needs to calculate $\Gamma_{k=M}$, where $M$ is the scale at which a perturbative description becomes possible. This procedure has been performed and carefully checked in QCD, where we flow from an asymptotically free theory of quarks and hadrons in the ultraviolet to chiral perturbation theory and low energy effective models in the infrared. 

First steps towards using the FRG to derive the effective action of quantum gravity and matter systems have been taken in \cite{Codello:2015oqa,Ohta:2020bsc}. Such calculations overlap significantly with 
EFT calculations.

Investigations in low energy effective theories are typically based on the Wilsonian action $S_{\textrm{eff},\Lambda}$ (regularized with a UV cutoff) both in QCD and in standard perturbative low energy EFT approach which is used in collider physics. The Wilsonian action is the generalized Legendre transform of the effective average action~\cite{Ellwanger:1993mw,Morris:1993qb} and obeys the Polchinski equation~\cite{Polchinski:1983gv}. So far, the Wilsonian action has been less used in asymptotic-safety investigations, but it could help to compare to results obtained from collider measurements of scattering observables for its closeness to low energy effective theories. These works can be based on recent proposals in \cite{deAlwis:2017ysy, Bonanno:2019ukb} for the flow of the Wilsonian action based on proper time regulator schemes~\cite{Bonanno:2019ukb}. 

The choice of truncation used for the effective average action down to the scale $M$ might be crucial to correctly encode  the various consequences of the UV fixed point, both in the matter and gravity sector.
$\Gamma_{k=M}$ or $S_{\textrm{eff},\Lambda=M}$ provide the initial condition for a subsequent perturbative
calculation at one or two loops; of course, also RG schemes would need to
  be matched for precision calculations. The perturbative part of the RG evolution gives rise to the non-localities in $\Gamma_{k\rightarrow 0}$ 
and all IR effects which are necessary to include to correctly describe observables. In this way, the FRG and perturbation theory can be used concertedly in order to link the UV fixed point to observable physics in the IR (see also Sec.~\ref{sec:CorrCoupOb}), and the use of different RG equations (Wetterich and Polchinski) would offer non-trivial consistency checks.

\subsection{Effective vs. fundamental Asymptotic Safety}
\label{subsec:EFASvsFunAS}

 \emph{...where we discuss why an asymptotically safe fixed point could matter even if the deep UV of quantum gravity is described by a completely different theory.}\\

The RG fixed point underlying asymptotic safety features infinitely many
infrared attractive directions. Therefore, a fixed point can serve various
  purposes in different scenarios: 
 1) it can be the UV starting point of an RG trajectory, 2) it can be the IR endpoint of an RG trajectory, 3) it can generate an intermediate scaling regime at finite scales. The latter option can play a role in settings where a more ``fundamental'' description of quantum gravity holds at small distance scales, i.e., beyond  a finite momentum cutoff $k_{\rm UV}$. Indeed, for $k<k_{\rm UV}$, an effective description (with the metric as the effective gravitational field -- not necessarily in the sense of perturbative EFT) holds, i.e., we are in the theory space of asymptotically safe gravity. The more fundamental description provides the initial condition for the RG flow at $k_{\rm UV}$. If the initial condition satisfies a finite number of conditions related to the relevant directions of the fixed point, the flow will pass  close by the fixed point and exhibit an approximate scaling regime over a finite range of scales. The flow towards the deep IR will then closely resemble that of an actual fixed-point trajectory, resulting in essentially the same predictivity \cite{Held:2020kze}, see \cite{Percacci:2010af} for a general discussion and \cite{deAlwis:2019aud} for a discussion in the context of string theory. 

In this sense, an asymptotically safe fixed point can play a role in an EFT setup for gravity, and serve as a way to extend the regime of validity of the standard perturbative EFT framework.

\subsection{The structure of the vacuum}
\emph{...where we caution that the true ground state of gravity
  might not be a flat background, making the bridge to the EFT setting
  potentially more intricate. This question has so far only been addressed
  within a  severe approximation of the dynamics and degrees of freedom.}\\

The EFT approach to quantum gravity typically quantizes (small) fluctuations about a flat background. To link asymptotic safety to the EFT regime, one must therefore explore whether a flat background is a self-consistent choice, i.e., whether the flat background corresponds to the ground state of the theory.

To date the only explicit investigation of the vacuum structure of  asymptotically safe gravity based on the effective action $\Gamma_{k=0}$ has been performed within the conformally reduced $R+R^2$-approximation\footnote{In this approximation, only fluctuations of the conformal factor are taken into account. Quite surprisingly, this appears to suffice to generate an asymptotically safe fixed point in simple truncations \cite{Reuter:2008wj}, contrary to the expectation that the important degrees of freedom in gravity are the spin-2 ones.} and a layered structure of the effective spacetime has been found within this simple truncation (borrowing terminology from a vacuum model of Yang Mills theory, it has been termed ``lasagna vacuum'') \cite{Bonanno:2013dja}. Thereby the spatial modulation of the metric cures the notorious conformal factor instability generating a phase similar to those present also in higher-derivative low-dimensional condensed matter systems.
 While this proposed vacuum structure has only been found in a severe approximation of the dynamics and degrees of freedom, this can be read as a firm warning regarding all backgrounds that are not shown to be solutions of the effective field equations. They are of no physical relevance and might convey an incorrect general picture. In particular, one typically expects truncations to converge faster when the field configurations are expanded about the true ground state of the theory -- an expectation that can be tested within, e.g., the O($N$) model. 
On the other hand, it is crucial to remark that a spatially modulated ground state appears to be difficult to reconcile with stringent tests of Lorentz symmetry in the gravitational and the matter sector. Further, while the conformal approximation could suffice to capture the presence of a fixed point, it is to be expected that the inclusion of spin-2 modes will have a strong impact on such studies.  

Moreover, the importance of properly accounting for the ($k$-dependent) ground state in studies of the flow
 is emphasized in a recent background-independent re-analysis of the cosmological constant problem allegedly caused by quantum vacuum fluctuations. Paying careful attention to identifying the correct ground state, the often discussed naturalness problem disappears, see \cite{Pagani:2019vfm}.

Understanding the ground state of the theory at $k =0$ is important.
 It is  expected that since $\Lambda_k = \lambda_* k^2 \rightarrow \infty$ in the quantum regime governed by the Reuter fixed point, the self-consistent metrics (cf.~Sec.~\ref{sec:background}) $\gb^{\rm sc}_{k \rightarrow \infty} \propto k^{-2}$ will display increasing and ultimately diverging curvature. It is an open question how  this manifests itself at the level of $\Gamma_{k\rightarrow 0}$ and its effective field equations. Whether this is an unphysical effect and only present at large $k$ or whether it translates into a physical scale dependence is presumably important for questions of singularity resolution in black-hole spacetimes and the early universe. More generally,
accounting for true vacuum of the theory, with the help of the self-consistent background is important
 for a quantitatively precise exploration of the phenomenological implications of the quantum-gravity effects. 

\subsection{RG improvement}\label{RGimpro}
\emph{...where we critically review and discuss the procedure of RG improvement, discuss its interpretation as 
``quantum-gravity inspired" phenomenology, and caution regarding the quantitative reliability of this tool.}\\

Since the task of calculating the
effective action $\Gamma_{k \rightarrow 0}$, including its non-local contributions,
is an extremely challenging one, one may hope to extract qualitative information on the effects of quantum fluctuations by applying the procedure of
``RG improvement'' in gravity.
In Sect.VI.A, we have already defined what is meant by RG improvement in a perturbative context.
Proceeding in a similar way in a gravitational context,
it has been a common strategy to retain the dependence of some of the couplings, $G_k$ and $\Lambda_k$ say, on the RG scale $k$ and identify the latter with a geometrical quantity or momentum.
Based on such RG improvement ideas there is a substantial body of work investigating black-hole physics \cite{Bonanno:1998ye,Bonanno:2000ep,Bonanno:2006eu,Cai:2010zh,Reuter:2010xb,Falls:2010he,Becker:2012js,Falls:2012nd,Torres:2013cya,Litim:2013gga,Koch:2013owa,Kofinas:2015sna,Pawlowski:2018swz,Adeifeoba:2018ydh, Platania:2019kyx}, gravitational collapse \cite{Casadio:2010fw,Fayos:2011zza,Torres:2014gta,Torres:2014pea,Bonanno:2016dyv,Bonanno:2017zen,Bonanno:2017kta}, and cosmological scenarios \cite{Bonanno:2001xi,Reuter:2004nv,Reuter:2005kb,Reuter:2007de,Bonanno:2007wg,Hindmarsh:2011hx,Copeland:2013vva,Becker:2014jua,Bonanno:2015fga, Bonanno:2016dyv, Kofinas:2016lcz,Bonanno:2017gji, Tronconi:2017wps,Bonanno:2018gck, Platania:2019qvo,Platania:2020lqb} inspired by Asymptotic Safety. One might expect that this procedure could be justified in some cases where the external scale in question acts as an IR cutoff for fluctuations.

The ``improvement'' could be applied at different stages, for instance, at the level of the action or the field equations, or of the solution of the field equation. This freedom already implies that this procedure could lead to ambiguous results.
As an example,
we may consider the RG-improvement procedure based on the effective average action approximated by the Einstein-Hilbert action,
\begin{equation}\label{Gkans}
\Gamma_k = \frac{1}{16\pi G_k} \int d^4x \sqrt{g} [2 \Lambda_k - R] \, .
\end{equation}
Dimensional analysis implies that at the fixed point $G_k = g_* k^{-2}$, $\Lambda_k = \lambda_* k^2$. Identifying $k^2$ with the Ricci scalar and substituting the result back into \eqref{Gkans} leads to a higher-derivative $R^2$ action. 
This is precisely what one would expect for the fixed-point action for an $f(R)$ theory
in the large $R$ limit. Indeed RG improving any $f(R)$ theory in the same way results in an $R^2$ action, as expected from classical scale invariance. 
Thus the scale identification generates interactions that have a natural place in the effective action \eqref{Geff}. 
However, this can lead at most to qualitative insights,
as is made clear, for example, by the fact that even the simple identification
$k^2\sim R$ can only be made up to some arbitrary numerical factor.

To understand better whether an RG improvement is justified,
let us consider some classic QFT examples, and contrast
them with their gravitational counterparts.
The Uehling potential in QED is probably the paradigmatic example: the correct form of the one-loop potential between two point charges can be obtained by inserting the one-loop form of the running coupling in place of the classical coupling and identifying the RG scale with the Fourier momentum of the static potential between the point charges. Conversely, one can read off the screening nature of the QED coupling from the one-loop effective action. 
Similarly, the Coleman-Weinberg effective potential is obtained,
in a classically scale-invariant theory, by replacing the
classical quartic scalar coupling by its one-loop
counterpart, evaluated at a renormalization scale
$k\sim\phi$. This is justified, insofar as the classical
VEV of the scalar is the only scale in the problem.
Similar considerations have also been applied
to non-Abelian gauge theories
\cite{migdal73,matinyan78,pagels78}.

Coming closer to gravity, a recent example in curved space where RG improvement works, 
is the case of interacting conformally coupled fields in de Sitter spacetime. 
A  correlator  evaluated at the fixed point can be related to a CFT correlator in flat space by a Weyl transformation. 
Then, the late time power-like behavior of correlators can be 
obtained as a resummation of secular terms controlled by the anomalous dimensions in flat space, with an RG improvement at the renormalization scale $\mu=H$~\cite{Green:2020txs}, where the Hubble scale $H$ of the de Sitter background is the only non-trivial scale in the problem.

Even more relevant for us, the running of $G$ and the quantum corrections to the Newtonian potential due to a scalar field loop have been compared in \cite{Dalvit:1994gf}.
They find that in general the RG improvement gives the expected qualitative behavior, and also reproduces the correct numerical coefficients for minimal coupling ($\xi=0$) or conformal coupling 
($\xi=1/6$).

The reason why all these examples work (at the quantitative level) is the logarithmic running of the coupling.
It is particularly instructive to compare the Uehling potential with 
the analogous calculation in gravity.
In the calculation of \cite{Dalvit:1994gf},
the running of $G$ is logarithmic and proportional to the mass
of the scalar field. This gives a result that is in agreement
with the quantum correction to the potential.
On the other hand, if one extracts the (quadratic) running of $G$ from the FRG, and tries to derive the analog of the Uehling
potential from there, one gets a term with the opposite sign
of the quantum correction calculated in EFT \cite{Donoghue:1993eb}.
This is a clear failure of the RG improvement:
the EFT calculation gives a screening contribution,
whereas the FRG seems to give an antiscreening one,
as required by asymptotic safety. 
The situation has been clarified in part in \cite{Becker:2014qya}: due to the use of the background field method,
there are different ways of defining Newton's coupling
that have different types of behavior at low energy (where the EFT result holds)
and at high energy, where one is assumed to approach a fixed point.
However, this leads us back to the issue of the shift Ward
identities, cf.~Sec.~\ref{sec:background} that, as discussed earlier, does not currently have a
satisfactory solution.

In conclusion, physical quantum effects in an asymptotically safe theory 
have to be calculated,
as in any other QFT, from the effective action, 
where all fluctuations have been integrated out.
We stress that the results one obtains from the
RG improvement,
e.g., for black holes or the early universe, 
cannot be viewed as actual derivations from a 
fundamental theory of quantum gravity,
but should still be viewed as ``quantum-gravity-inspired models'',
providing  qualitatively sensible, though not necessarily precise, answers 
in some cases where there is a clearly identifiable 
single scale in the problem.

\section{Scale symmetry and conformal symmetry}
\label{sec:ScaleA+Conformal}

\subsection{The RG as scale anomaly}
\emph{...where we clarify the meaning of scale symmetry in the context of asymptotic safety.}\\

A point that  tends to generate confusion 
concerns the interpretation of the RG flow
as an anomalous breaking of scale invariance.
It may seem puzzling that the asymptotic safety program
claims (quantum) scale invariance even though $\Gamma_k$ contains
dimensionful couplings.
The goal of this subsection is to clarify this point. We follow \cite{Morris:2018zgy},
see also \cite{Wetterich:2019qzx}, section 6.9.
\footnote{Here we discuss global scale invariance.
It has been shown that local scale invariance
can be maintained in the RG flow provided a dilaton field is present
\cite{Percacci:2011uf,Codello:2012sn,Pagani:2013fca}.}

Consider a perturbatively renormalizable QFT, 
with an interaction term $u\,\cO\equiv u\int d^4x\cL$, where $\cL$ 
is a dimension-four operator and $u$ a dimensionless coupling.
If there is no mass term, the theory is scale invariant
under the standard realization of scale transformations
which act on the fields but not on the couplings.
In the quantum theory, however, 
scale invariance is broken by the beta function
\begin{equation}
\delta_\epsilon \Gamma=-\cA(\epsilon)
\sim -\epsilon \beta_u\,\cO\ .
\label{anom}
\end{equation}
Here $\epsilon$ is the infinitesimal parameter generating the transformation, $\delta_\epsilon g_{\mu\nu} = 2 \epsilon g_{\mu\nu}$, etc.\ and $\cA(\epsilon)$ is the trace anomaly which can be formally 
seen as due to non-invariance of the functional integration measure. 
At a fixed point $\beta_u = 0$ and scale invariance is recovered.

Eq.\ \eqref{anom} can be generalized to the Wilsonian RG.
In this case there is an additional term coming from the presence of an explicit 
momentum cutoff which is given by the ``beta functional'' defined in (\ref{FRGE}):
\begin{equation}
\delta_\epsilon \Gamma_k=-\cA(\epsilon)
+\epsilon k\partial_k\Gamma_k\ .
\label{anomw}
\end{equation}
For the effective average action given in  
(\ref{eq:operatorexpansion})
one finds that the anomaly is given by \cite{Morris:2018zgy}
\begin{equation}\label{anom3}
\cA
\sim \epsilon \sum_i\beta_{u_i}\,k^{d_i} \, \cO_i\ ,
\end{equation}
where $d_i$ is the canonical mass dimension of $\cO_i$. 
Again $\cA$ vanishes at a fixed point. Nevertheless, 
the standard realization of scale invariance, acting on fields only, is 
broken due to the extra term in \eqref{anomw}
\begin{equation}
\delta_\epsilon\Gamma_k
\sim \epsilon \sum_i \, d_i \, \bar{u}_i \, \cO_i\ .
\end{equation}
There is however an alternative realization of scale invariance 
acting on both the fields and the cutoff. Here 
the transformation of the fields remaining unaltered
$\hat{\delta}_\epsilon g_{\mu\nu} = 2 \epsilon g_{\mu\nu}$, etc.\ while
the cutoff transforms as $\hat\delta_\epsilon k=-\epsilon k$.
Under this alternative realization,
\begin{equation}
\hat\delta_\epsilon\Gamma_k
\sim -\cA(\epsilon)\ ,
\end{equation}
which vanishes at a fixed point. 

In conclusion, we see that in a ``Wilsonian'' formulation of the RG,
quantum scale invariance is realized at a fixed point, albeit with respect to a different implementation of rescalings than the one generally used in particle physics.

\subsection{Black hole entropy}
\emph{...where we discuss an argument against a QFT for gravity based on black-hole entropy and point out where assumptions are being made which require further investigation.}\\

Refs.\ \cite{Aharony:1998tt,Shomer:2007vq} presented a chain of arguments indicating that a quantum-field theoretic description of gravity in four dimensions cannot be UV complete. In short, this chain proceeds along the following lines. First, it is assumed that, at high energies, the density of states in quantum gravity is dominated by black holes, which also goes by the name of ``asymptotic darkness''. Black hole thermodynamics, building on quantum field theory on a curved background, implies that the leading term in the entropy $S$ of the black hole is proportional to the area $A$ of its horizon. For a $d$-dimensional Schwarzschild black hole
\begin{equation}\label{arealaw}
	S \propto A \propto M^{\frac{d-2}{d-3}}
\end{equation}
where $M$ is the ADM mass of the black hole. Identifying $M$ with a typical energy scale $E$, the asymptotic darkness hypothesis then suggests that the number of states available at high energy should scale as
\begin{equation}\label{bhent}
S^{\rm BH} \propto E^{\frac{d-2}{d-3}}
\end{equation}
In four dimensions this implies that $S^{\rm BH} \propto E^2$. On the other hand, the degrees of freedom of a conformal field theory (CFT) living on a $d$-dimensional Minkowski space follow the scaling law
\begin{equation}\label{cftent}
S^{\rm CFT} \propto E^{\frac{d-1}{d}}
\end{equation}
which in four dimensions becomes $S^{\rm CFT} \propto E^{\frac{3}{4}}$. The mismatch between the density of available states \eqref{bhent} and \eqref{cftent} is then taken as an indication that the high-energy completion of four-dimensional gravity cannot be given by a conformal field theory. 

We now critically review the assumptions entering into this chain of arguments: \\
{\bf 1}) Scales involved in the problem: \\ Seeing quantum-gravity effects in scattering events requires going to \emph{large energies and small impact parameters} relative to the Planck scale. This is not the same as considering just trans-Planckian energies: the energy involved in the merger of two astrophysical black holes clearly exceeds the Planck mass $m_{\rm Pl} \approx 10^{-5}g$ by many orders of magnitude. Nevertheless, classical general relativity provides a very accurate description of these events, for which the impact parameter is large compared to the Planck length.\\
{\bf 2}) The asymptotic darkness hypothesis: \\
The idea of asymptotic darkness relies on the hoop conjecture~\cite{Thorne:1972ji} which states that scattering at sufficiently high energy  results in black-hole formation. While numerical simulations confirm this expectation in classical gravity \cite{Choptuik:2009ww,East:2012mb}, a corresponding study in the quantum case is lacking, see also the discussion in \cite{Basu:2010nf}. When phrased in terms of the effective action \eqref{Geff}, it is expected that the form-factors $W(\Delta)$ (or, more generally, the 1PI vertices) will play a central role in correctly describing scattering processes at trans-Planckian scales. Currently, little is known about these effects though, and it is an open question whether or not Planckian scattering in asymptotically safe gravity does or does not lead to black-hole formation. 
In \cite{Dvali:2010jz}, it has been proposed that black-hole formation in Planckian scattering is a key property of gravity that allows the theory to self-unitarize (classicalisation). Whether this has anything to do with asymptotic safety is an open question.
See \cite{Percacci:2012mx,Aydemir:2012nz} for related discussions in the context of non-linear sigma models.
\\
{\bf 3}) Corrections to the entropy formula: \\
The semi-classical area law \eqref{arealaw} is a good approximation for large black holes. It receives further corrections from quantum gravity though. Logarithmic corrections were determined in \cite{Kaul:2000kf}, indicating that
\begin{equation}
S = \frac{A}{4G} - \frac{3}{2} \log\left(\frac{A}{4G}\right) + \cdots 
\end{equation} 
Clearly, these corrections become increasingly important for small (i.e., near-Planckian) black holes, see, e.g., \cite{Kaul:2000kf,Carlip:2000nv,Engle:2009vc,El-Menoufi:2015cqw}. Thus, it is a priori unclear if the simple scaling law \eqref{arealaw} is applicable in the quantum gravity regime.\\
{\bf 4)} Dimensional reduction of the momentum space: \\
A critical point in extending scaling arguments to quantum gravity is the identification of the correct notion of dimensionality which actually controls the scaling laws. While in flat Minkowski space there is just the dimension of spacetime $d$, fluctuating spacetimes are typically characterized by a whole set of ``generalized dimensions'' (spectral dimension, Hausdorff dimension, etc.) which do not necessarily agree. In particular, a rather universal result about quantum gravity \cite{Carlip:2017eud,Carlip:2019onx} indicates that the dimension of the theory's momentum space (spectral dimension) undergoes a dimensional reduction to $d_s = 2$ at energies above the Planck scale. In \cite{Doboszewski:2017gim}, it was argued that such a mechanism could constitute a potential way to reconcile the semi-classical scaling in gravity with the scaling of states in the conformal field theory. In order to make such proposals robust, it is important to identify the proper notion of dimensionality which controls the scaling of the quantity of interest. In the context of black hole thermodynamics, it has been suggested that this could be achieved with the ``Unruh dimension''  \cite{Alkofer:2016utc} governing the scaling laws in the black-hole evaporation process. \\
{\bf 5}) Entropy of asymptotically safe black holes: \\
The entropy of black holes in asymptotic safety has been investigated in \cite{Bonanno:2000ep,Reuter:2010xb,Falls:2010he,Koch:2013owa} based on RG improvement techniques (the cautionary remarks regarding RG improvement from Sec.~\ref{RGimpro} apply in this case). One outcome of this investigation was that the entropy of Planck-size black holes follows the Cardy-Verlinde formula \cite{Koch:2013owa} indicating compatibility with a conformal field theory description. Concerning macroscopic black holes, the semi-classical result for the black-hole entropy can presumably be understood entirely in terms of the entanglement entropy of matter fields living on the black-hole background geometry \cite{Sorkin:2014kta}, see \cite{Solodukhin:2011gn} for a comprehensive review, and  \cite{Jacobson:2012ek,Pagani:2018mke} for discussions in the context of the FRG and asymptotic safety.

In conclusion, combining semi-classical arguments based on the asymptotic darkness hypothesis and conformal field theory in flat space gives rise to results in tension with the asymptotic-safety conjecture. It is clear that much more work is needed in order to actually show that these arguments also apply in the framework of quantum gravity.

\section{Unitarity}
\label{sec:Unitarity}

\subsection{General remarks}\label{sec:unitarity_general}
\emph{...where we point out that the concept of unitarity in quantum gravity is way more subtle than for a quantum field theory on flat Minkowski space.}\\

Conservation of probabilities is a cornerstone of quantum mechanics. For a QFT in a flat Lorentzian background, this feature is reflected by the $S$ matrix, connecting  the initial state and the final state of a physical system, being unitary. Starting from a QFT defined on a Euclidean signature spacetime the Osterwalder-Schrader axioms \cite{Osterwalder:1973dx,Osterwalder:1974tc}, including the requirement of reflection positivity, guarantee that the theory has an analytic continuation to a unitary QFT.

Notably, it is highly non-trivial to generalize the concept of a unitary $S$ matrix to more general backgrounds \cite{Bousso:2004tv} or to the gravitational interactions \cite{Giddings:2009gj, Giddings:2011xs}. Examples for such generalizations are the local $S$ matrix in de Sitter space studied in \cite{Marolf:2012kh} or the one recently constructed in \cite{David:2019mos}.
  
Along a different line, the existence of unphysical modes such as tachyons, negative norm states, etc., in a given background $\bar{g}_{\mu\nu}$ does not automatically signal the inconsistency of the theory. It may just indicate the instability of this particular background\footnote{This is a well-known situation, for instance, in scalar theories, where an expansion about a saddle point of the potential leads to tachyonic instabilities, but does of course not signal an inconsistency of the theory. For instance, in inflationary scenarios these instabilities are key to the resulting physics.}. As an example, \cite{Bonanno:2013dja} highlights how a non-standard background removes the conformal-mode problem in the Euclidean path-integral. From a phenomenological point of view, a minimal requirement is to impose that cosmologically relevant backgrounds of Friedman-Lemaitre-Robertson-Walker-type are stable on cosmic time-scales.

An important indicator that asymptotically safe gravity could indeed be unitary comes from the causal dynamical triangulations (CDT) program. Here one finds that the (two-step) transfer matrix connecting spacial slices at different time-steps is self-adjoint and bounded  \cite{Ambjorn:2000dv,Ambjorn:2001cv}, indicating that it satisfies the requirement of reflection positivity. Since the analytic continuation to Lorentzian signature is well-defined in CDT, the resulting Hamiltonian in the Lorentzian setting is self-adjoint.
Under the preconditions that this feature survives in the continuum limit and that CDT indeed probes the Reuter fixed point, this indeed points towards Asymptotic Safety being a unitary theory. 

These limitations should be kept in mind when discussing unitarity in a background-independent, quantum gravitational setting. \\

\subsection{Flat-space propagators}\label{sec:FlatProp}
\emph{...where we review Ostrogradsky's theorem and its loopholes.}\\

With the above cautionary remarks in mind, let us discuss the gravitational propagator on a flat background. In the presence of a finite number of higher-derivative terms, a partial-fraction decomposition of the propagator reveals the presence of additional modes. For example, a propagator derived from a four-derivative theory yields
\begin{equation}\label{eq:ghostmode}
\frac{1}{p^2(p^2+m^2)} = \frac{1}{m^2} \left( \frac{1}{p^2} - \frac{1}{p^2 + m^2} \right) \, .
\end{equation}
The terms in the partial-fraction decomposition come with alternating signs with the modes associated to the negative residues corresponding to ghosts.  In the case of physical fields related to asymptotic states, this violates reflection positivity \cite{Arici:2017whq}. The latter signals the violation of unitarity in the Lorentzian theory and is related to a spectral function with negative parts. The violation of unitarity is already present at the classical level where it corresponds to an instability of the theory according to Ostrogradsky's theorem. Any 
non-degenerate Hamiltonian with higher time derivatives of finite order unavoidably features such an instability, see, e.g, the pedagogical discussion in \cite{Woodard:2015zca}. This directly implies that truncations to finite order in momenta generically feature truncation-induced instabilities and are not suitable to investigate the unitarity of the theory.

There are three prominent ways to avoid the Ostrogradsky instability: \\
{\bf 1)} Propagators consisting of an entire function with a single zero at vanishing momentum may avoid the occurrence of negative residues. This is the path taken by non-local ghost-free gravity \cite{Biswas:2005qr,Modesto:2011kw,Biswas:2011ar,Modesto:2017sdr}. At the classical level, the well-posedness of the corresponding initial-value problem has been discussed in \cite{Barnaby:2007ve}. \\
{\bf 2)} One can give up Lorentz invariance, introducing higher-order spatial derivative terms while keeping two time-derivatives. This is the idea underlying Ho\v{r}ava-Lifshitz gravity \cite{Horava:2009uw} which, by construction, is a power-counting renormalizable, unitary theory of gravity. \\
{\bf 3)} Accept that Nature allows for the violation of causality at microscopic levels \cite{Holdom:2015kbf,Anselmi:2018ibi,Donoghue:2019fcb,Donoghue:2019ecz}. In this case, the degrees of freedom associated with negative residues are interpreted as ``particles propagating backward in time''. If these particles are sufficiently heavy this may not leave an experimentally detectable trace.

We stress that in any case unitarity should be assessed based on the propagators derived from the effective action $\Gamma_{k=0}$. Propagators derived from the effective average action $\Gamma_k$ at intermediate $k$ may feature artificial poles. Under the flow in $k$, the mass of a ghost might diverge so that the corresponding degrees of freedom decouple, see \cite{Becker:2017tcx}.

 \subsection{Spectral function of the graviton}

\emph{...where we discuss the consequences of potential negative spectral weights of the graviton.}\\
 
The ghost mode discussed in the last Sect.~\ref{sec:FlatProp} is but one example for 
a spectral function that has negative spectral weights: Evidently, the second term in parenthesis in eq.~\eqref{eq:ghostmode} leads to a $\delta$-function with negative prefactor in the spectral function of the graviton. In asymptotically safe gravity the graviton propagator is a general function of momentum. Consequently the spectral function more generally may simply have negative parts. 

To begin with, negative spectral weights are a well-known feature of the gluon spectral function in Yang-Mills theory: upon the assumption of a spectral representation of the gluon, it can be shown that its total spectral sum is vanishing due to the Oehme-Zimmermann superconvergence relation \cite{Oehme:1979bj, Oehme:1990kd}. This relation already
implies that in the asymptotically free regime of Yang-Mills
theory, the spectral function of the gluon is negative for large spectral values. Indeed, the analytic form for large spectral values can be computed within perturbation theory. More recently it could also be shown by similar arguments that the spectral function is also negative for small spectral values \cite{Cyrol:2018xeq}.

These investigations highlight the fact that even the existence of spectral representations for gauge fields with non-linear gauge symmetries is an open issue. This is tied to the fact that these fields are not directly related to asymptotic states even in regimes where they heuristically can be interpreted as particles. In QCD this is manifest in gluon jets at colliders. In the context of gravity, this feature is intrinsic to the proposal made in point 3) of the previous subsection: owing to their large mass, the states associated with the negative residue terms do not correspond to asymptotic states, see \cite{Maas:2019eux} for a recent discussion.

In summary, even if the spectral representation of a gauge fields exists, it very well can -- and in the case of the gluon must -- contain negative parts. Evidently, this adds significantly to the already discussed intricacies of discussing  unitarity and the interpretation of positivity violations in quantum gravity: negative spectral weights may be present without spoiling unitarity but clearly their presence casts doubts on unitarity. This situation asks for the investigation of the spectral representation of correlations of well-defined diffeomorphism-invariant observables, see Sect.~\ref{sec:Observables}.

\subsection{Interpretation of potential ghost modes}

\emph{...where we refer back to the concept of effective asymptotic safety discussed in Sec.~\ref{subsec:EFASvsFunAS} and discuss the interpretation of the masses of unstable graviton modes in this context.}\\

Future studies of unitarity may reveal that asymptotic safety features {\it physical} ghost modes and hence is not a unitary fundamental QFT. Even in this case, an asymptotically safe fixed point can still play a role within the setting described in Sec.~\ref{subsec:EFASvsFunAS}, and serve as an extension of the EFT regime for gravity. Then, the asymptotically safe description in this setting could inherit  unitarity from the more ``fundamental" description. In particular, the asymptotically safe setting can in this case exhibit unstable modes, with masses $m>k_{\rm UV}$ -- these signal the need for a more ``fundamental" description. Conversely, the masses of ghost modes can be used as an estimate for the scale of new physics. \\

\subsection{Remarks}

In summary, it is currently unclear whether or not asymptotically safe gravity is unitary, it shares with other approaches to quantum gravity.  
The question combines both conceptual and technical challenges in quantum gravity: there is the conceptual question of the complex structure of correlation functions in the presence of a dynamical metric field, as well as the necessity of \textit{non-perturbative} numerical computations in Lorentzian signature. 
As already emphasized in Sect. \ref{sec:Observables}, cross-checks between quantum-gravity approaches and the concerted use of more than one method are called for.

\section{Lorentzian nature of quantum gravity}
\label{sec:LorentzianGravity}
\emph{...where we highlight the expected fundamental difference between Lorentzian and Euclidean quantum gravity and explain why the flow equation is typically set up in a Euclidean setting.}\\

Hitherto, the bulk of the Asymptotic Safety literature employs background spacetimes carrying Euclidean signature metrics. This brings two technical advantages: Firstly, Euclidean signature entails that the squared momentum of the fluctuation fields is positive semi-definite. Thus it is straightforward to define the ``direction of the RG flow'', first integrating out fluctuations with a large squared momentum before successively moving towards lower values. Secondly, the regulated propagators do not exhibit poles, as the particle cannot go on shell. 

For a QFT defined on flat Euclidean space $\mathbb{R}^d$, one can carry out the computation and analytically continue the results to Lorentzian signature by a Wick rotation. In the context of quantum gravity, including Asymptotic Safety, this strategy is very challenging for several reasons listed below, part of which has been already discussed in detail in Sect.~\ref{sec:Unitarity}. 
\begin{enumerate}[leftmargin=0.5cm]
	\item A generic background metric may not admit a (global) Killing vector which lends itself to an analytic continuation to a well-defined Lorentzian time direction \cite{Baldazzi:2018mtl}. 
	\item The complex structure of the full graviton propagators may obstruct the simple analytic continuation of the Euclidean propagator, for example there may very well be cuts touching the Euclidean axis. Within the FRG this is complicated further as a momentum regularization either breaks the underlying (global) spacetime symmetry or leads to additional poles and/or cuts \cite{Pawlowski:2015mia}. 
		There has been much progress in the past years on this in standard QFTs, see e.g., \cite{Floerchinger:2011sc, Kamikado:2013sia, Pawlowski:2015mia, Yokota:2016tip, Wang:2017vis, Tripolt:2018qvi},  but the extension to asymptotically safe gravity has not been put forward yet. 

	\item At the structural level, there are solid arguments to expect that the effective actions obtained from integrating out fluctuations in a Lorentzian and Euclidean signature setting will be different. Firstly, the space of metrics of the two settings comes with different topological properties: while all Euclidean metrics can be connected by geodesics (defined with respect to a suitable connection) this property does not hold in the Lorentzian case \cite{Demmel:2015zfa}. Secondly, the heat kernels for differential operators constructed from a Euclidean and Lorentzian signature metric differ by non-local terms \cite{Vassilevich:2003xt}. While the latter do not affect the singularity structure of the heat kernel underlying perturbative renormalization, they may lead to differences in $\Gamma$.
\end{enumerate}

A way to address the first point comes from studying Asymptotic Safety in the Arnowitt-Deser-Misner (ADM) formalism. In this case, spacetime has a built-in foliation structure which defines a natural time direction. A first investigation of Asymptotic Safety in this framework has been performed in \cite{Manrique:2011jc} and further developed in a series of works \cite{Rechenberger:2012dt,Contillo:2013fua,Biemans:2016rvp,Biemans:2017zca,Houthoff:2017oam,Knorr:2018fdu,Eichhorn:2019ybe,Nagy_2019}. This provided first-hand indications that the asymptotic-safety mechanism remains operative for Lorentzian signature metrics as well, at least within very small truncations. At this stage the computations in the Lorentzian signature framework have not reached a level of sophistication where the structural differences outlined in point 3) can be resolved. In general, the systematic development of the FRG applicable to Lorentzian signature spacetimes is a research area to be developed in the future.

This point could in the future become another example for the progress that can become possible if tools and concepts from various quantum-gravity approaches are brought together. Specifically, causal set theory (see \cite{Surya:2019ndm} for a review),
at least when restricted to so-called ``sprinklings'',
can be viewed as a discretization of the Lorentzian path integral over geometries. See also \cite{Eichhorn:2019xav}. This motivates the search for a universal continuum limit, linked to a second-order phase transition in the phase diagram for causal sets. Monte Carlo studies of the phase diagram for restricted configuration spaces in low dimensionalities can be found in \cite{Surya:2011du,Glaser:2017sbe,Glaser:2018jss,Cunningham:2019rob}.

\acknowledgements
We are indebted to A.~Codello, J.~Donoghue, K.~Falls, A. Held, B.~Knorr, T.~R.~Morris, D.~F.~Litim,  R.~Loll, A.~Pereira, A.~Platania, N.~Ohta, M.~Reichert, C.~ Ripken, C.~Wetterich, and O.~Zanusso for valuable discussions and comments on this manuscript.
A.~E.~is supported by the DFG (Deutsche Forschungsgemeinschaft) under grant no.~Ei/1037-1, by a research grant (29405) from VILLUM FONDEN and by a visiting fellowship of the Perimeter Institute for Theoretical Physics. H.~G.~is supported by the DFG under grant no.~398579334 (Gi328/9-1). The work of F.S.\ is supported by the Netherlands Organisation  for  Scientific  Research  (NWO)  within  the  Foundation  for  Fundamental Research on Matter (FOM) grant 13VP12. J.~M.~P.~is supported by the DFG Collaborative Research Centre SFB 1225 (ISOQUANT) as well as by the DFG under
Germany's Excellence Strategy EXC - 2181/1 - 390900948 (the
Heidelberg Excellence Cluster STRUCTURES).

\bibliographystyle{apsrev4-1}
\bibliography{refs.bib}
\end{document}